\documentclass[preprint,11pt]{elsarticle}
\usepackage[english]{babel}
\usepackage[T1]{fontenc}
\usepackage{diagbox}
\usepackage{graphicx,booktabs}
\usepackage{caption}
\usepackage{subcaption}
\usepackage{setspace}
\usepackage{tikz}
\usetikzlibrary{arrows.meta, positioning, shapes.geometric}
\tikzset{
  block/.style={
    rectangle, rounded corners, draw,
    align=center,
    minimum height=0.9cm,
    text width=4cm,
    font=\small
  },
  arrow/.style={-Latex}
}

\usepackage{atbegshi}% http://ctan.org/pkg/atbegshi
\usepackage[T5]{fontenc}
\usepackage{natbib}
\usepackage{lineno,hyperref}
\usepackage{enumitem}
\usepackage{mathtools}
\usepackage[mathscr]{eucal}
\usepackage{dsfont}
\usepackage{array}
\usepackage{afterpage}
\usepackage{pdflscape}
\usepackage{caption}
\usepackage{subcaption}
\usepackage{graphicx}%
\usepackage{hhline, multirow}%
\usepackage{amsmath}
\usepackage{amssymb}
\usepackage{amsfonts}
\usepackage{amsthm}
\usepackage{multicol}
\usepackage{mathrsfs}%
\usepackage[title]{appendix}%
\usepackage{enumitem}
\usepackage{lscape}
\usepackage{adjustbox}
\usepackage{textcomp}%
\usepackage{manyfoot}%
\usepackage{booktabs}
\usepackage{tabularx}
\usepackage{array}
\usepackage{geometry}
\usepackage{algorithm}%
\usepackage{algorithmicx}%
\usepackage{algpseudocode}%
\usepackage{listings}%
\usepackage[hang,flushmargin]{footmisc} 
\usepackage{siunitx}
\usepackage{float}
\usepackage{siunitx} % For aligning decimals
\usepackage{multirow} % For multirow cells
\usepackage[table]{xcolor} % For cell coloring
\renewcommand{\arraystretch}{1.1} % slightly tighter spacing

\makeatletter
\def\ps@pprintTitle{%
   \let\@oddhead\@empty
   \let\@evenhead\@empty
   \let\@oddfoot\@empty
   \let\@evenfoot\@empty}
\makeatother

\usepackage[table]{xcolor}  % Optional: for better visuals

\usepackage{calrsfs}
\usepackage[
    left = \flqq{},% 
    right = \frqq{},% 
    leftsub = \flq{},% 
    rightsub = \frq{} %
]{dirtytalk}
\newcolumntype{R}[1]{>{\raggedleft\arraybackslash}p{#1}}
\newcolumntype{L}[1]{>{\raggedright\arraybackslash}p{#1}}
\newcolumntype{P}[1]{>{\centering\arraybackslash}p{#1}}
\newtheorem{remark}{Remark}
\newtheorem{defn}{Definition}
\newtheorem{prop}{Proposition}

\DeclareMathAlphabet{\pazccal}{OMS}{zplm}{m}{n}

\setcitestyle{square}

\begin{document}
\begin{frontmatter}
%\title{When Volatility is Fair? An Efficiency-Consistent \\Definition of Financial Risk}
%\title{Reconceptualizing Financial Risk:\\ Fair Volatility and (In)Efficient Markets}
%\title{Fair Volatility: A Framework for Reconceptualizing Financial Risk}
\title{When is Volatility Fair? \\Hölder Regularity and Financial Risk}
%\title{Financial risk and \textit{fair} volatility}
%\affiliation[first]{organization={Sapienza University of Rome},%Department and Organization
%            addressline={Via del Castro Laurenziano, 9}, 
%            city={Rome},
%            postcode={00161}, 
%            country={Italy}}
            %\cortext[cor1]{Corresponding author, sergio.bianchi@uniroma1.it}
%\affiliation[fourth]{organization={University of Cassino and Southern Lazio},%Department and Organization
%            addressline={Via S. Angelo}, 
%            city={Cassino},
%            postcode={03043}, 
%            country={Italy}}
\author{Sergio Bianchi and Daniele Angelini}%\corref{cor1}
%\author{Daniele Angelini}
%\author[first]{\\Massimiliano Frezza}
%\author[fourth]{Augusto Pianese}

%\author{Sergio Bianchi \\ \textit{Sapienza University of Rome, Via del Castro Laurenziano, 9, Rome, 00161, Italy}\\sergio.bianchi@uniroma1.it}
\affiliation{Sapienza University of Rome, Via del Castro Laurenziano, 9, Rome, 00161, Italy} % Primary affiliation
%\email{sergio.bianchi@uniroma1.it}

\begin{abstract}
%Volatility's dominance as a risk measure of financial markets reflects historical convenience rather than theoretical superiority. Its mathematical tractability and ease of calculation have privileged it over more nuanced risk measures that better capture the complexities of financial markets. Moving beyond volatility requires developing risk measures that account for path-dependence and the dynamic nature of market efficiency. 
%In response, we propose the pointwise regularity (Hurst–Hölder exponent) as a complementary risk metric. This measure captures deviations from martingale behavior, offering a deeper understanding of market inefficiencies and the dynamics of reversion to equilibrium. By quantifying not just the magnitude but the nature of randomness, this framework reconciles insights from both efficient market theory and behavioral finance.
\footnotesize{Positing that the real nature of financial risk is not \emph{variability} itself, but rather \emph{irregularity} -- i.e. \emph{unpredictability} -- of price dynamics, we introduce a framework for measuring financial risk based on the local regularity of log-returns, captured by the time-varying Hurst–H\"{o}lder exponent. The paradigm shift resulting from disentangling variability and irregularity naturally leads to  define the notion of \emph{fair volatility}, i.e. the volatility level consistent with efficient market and martingale dynamics. In this view, fair volatility describes the maximum level of financial risk, which does not coincide with the higher volatility.
Within the very large class of Multifractional Processes with Random Exponent (MPRE), we establish an analytical relation between regularity and scale of price increments. Applying existing estimators, we compare MPRE-implied volatility with realized volatility across fourteen international equity indices. The results reveal coherent volatility patterns across markets and time, and highlight phases in which temporary inefficiencies occur that are corrected by types of opposite behavioral schemes.
}
\end{abstract}



\begin{keyword}
%% keywords here, in the form: keyword \sep keyword, up to a maximum of 6 keywords
\footnotesize{Stochastic Volatility \sep Financial risk \sep Hölder regularity \sep Hurst exponent \sep Multifractional Processes with Random Exponent}
\end{keyword}
\end{frontmatter}

%\tableofcontents


\section*{Introduction and motivation}
%**** SCITE
Volatility is widely regarded as a preeminent proxy for financial risk in both academic research and investment practice. This convention largely stems from Markowitz's Modern Portfolio Theory, which established variance minimization as the mathematical foundation in portfolio optimization \citep{LS2018}. In this work, we argue that the commonly assumed direct relationship between volatility and risk -- namely that higher volatility implies higher risk -- holds only under the assumption that the price process behaves as a local martingale \citep{RaoPeng2022,shiller1981use}, a condition typically justified by the posited ubiquitous efficiency of financial markets. However, this premise becomes misleading when more complex and realistic price dynamics are taken into account.% and aligns with broader concerns in risk analysis regarding the interpretability and consistency of commonly used risk indicators \citep{Borgonovo2025}. 

Actually, several issues challenge the robustness of volatility when the (sub)martingale paradigm is relaxed or when derivative assets are considered.  These issues concern: (a) the insensitivity to the temporal structure of returns and to forms of non-stationarity affecting data \citep{CV2003}; (b) the different meaning that volatility takes in portfolios with certain derivatives (no longer a measure of loss risk but an opportunity) \citep{Benbachiretal2025}; (c) the relative nature of volatility, which obscures what an appropriate level actually is \citep{Bisewskietal2021}.

\subsection*{a) The Temporal Blindness of Volatility}
The principal limitation of volatility as a risk indicator is its invariance to the temporal structure of returns. Sequences with identical unconditional volatility may display profoundly different temporal patterns, but this basic consideration is overlooked by the Efficient Market Hypothesis (EMH) paradigm because of the assumption that prices adjust rapidly to information and increments are therefore approximately independent \citep{F1970, Titan2015, Jangetal2019}. As a consequence, the distinction between \emph{return variability} and \emph{return predictability} is largely moot in the  semimartingale framework, the natural mathematical expression of EMH, because the predictable component of returns is excluded \emph{a priori}: conditional expectations of increments are null, and the only source of randomness is the martingale part. Consequently, volatility becomes the natural and essentially unique quantity available to characterise financial risk, as it estimates the quadratic variation when sampling intervals shrink. In such a world, variability exhausts the notion of risk precisely because 
predictability is ruled out by assumption. 

Although empirical evidence seems to support the independence of returns and confirm the EMH mechanism, the empirical finding of weak autocorrelation at daily horizons is not dispositive: non-stationarity may induce a “cancellation of dependence”, masking alternating regimes of persistence and antipersistence when returns are aggregated \citep{Anderson1994, Nielsen2006}.

A more explicit attention to temporal dependence emerged with the development of stochastic–volatility models designed to capture long-memory \citep{BaillieBollerslevMikkelsen1996, ComteRenault1998}, multiscale 
\citep{Corsi2009}, or rough persistence \citep{BayerFrizGatheral2016, Gatheral2018, Bennedsen2021}; in this regard, significant contributions have been made to both regime shift problems \citep{Grzelach2024} and efficient calibration of volatility surfaces \citep{Grzelak2015} under advanced stochastic and local volatility models. 
This literature shows that, even when returns remain conditionally unpredictable and therefore consistent with the martingale formulation of market efficiency, the second-order dynamics of prices can display persistent and economically meaningful temporal structure.

The evident limits of this standard environment motivate a broader framework in which risk measurement reflects not only the magnitude of fluctuations but also the degree of possible local predictability encoded in price dynamics. Extending volatility-based metrics to incorporate richer temporal features is therefore a natural conceptual reorientation when one moves beyond the strict semimartingale–EMH setting.

\subsection*{b) Derivative Strategies and the Volatility-Risk Disconnect}
The link between volatility and risk weakens further when evaluating strategies that incorporate options or other derivatives. For some strategies, increases in volatility may enhance expected returns rather than amplify risk \citep{CarrLee2009}. A canonical example is the long straddle -- combining a call and a put with identical strike -- whose payoff improves with large price movements in either direction. For market makers and volatility traders, volatility thus constitutes an opportunity set rather than a risk factor, as their exposure depends not on volatility per se but on its variations, including volatility-of-volatility and interactions across higher-order distributional features.

This underscores the strategy dependence of the volatility–risk relationship \citep{moreira2017volatility}, thereby challenging the view of volatility as a universal and objective metric of financial risk.\footnote{As Markowitz famously remarked: “\textit{We next consider the rule that the investor does (or should) consider expected return a desirable thing and variance of return an undesirable thing}.” \citep{Markowitz1952}}

\subsection*{c) The Missing Benchmark: What Volatility Should Efficient Markets Exhibit?}
Volatility lacks a natural benchmark: unlike asset prices, which are anchored by valuation principles such as discounted cash-flow models or arbitrage relationships, in an efficient market there is no theoretical criterion specifying what constitutes an “appropriate’’ or “normal’’ level of volatility. Because volatility is intrinsically variation-based, historical comparisons reveal only relative differences and do not indicate whether a given realisation is economically justified.

This indeterminacy stems from the fact that the continuous Brownian/Itô semimartingale $dX_t = \mu_t\,dt + \sigma_t\,dW_t$ can exhibit any level of local variance rate $\sigma_t$. Nonetheless, at any time $t$ its regularity, captured by the pointwise Hölder exponent, is always $\alpha_X(t)=\frac{1}{2}$, irrespective of the local variance rate. Quadratic variation encodes the amount of accumulated variance, whereas the Hölder exponent captures the scaling of increments. In this work, we exploit precisely this 
invariance of $\alpha_X(t)$ to deduce the level of volatility consistent with the semimartingale behavior so as to address the question ``when is volatility fair?’’—that is, when is a given volatility level consistent with semimartingale behaviour.

Absent such a link, efficient markets may display either low or high volatility, with no theoretical guidance for distinguishing normal fluctuations from those indicative of excessive risk or local disequilibria. Beyond limiting the regulatory usefulness of volatility, this constraint also obscures the fact that higher volatility need not imply greater financial risk. Decoupling predictability and variability helps clarifying that market efficiency and inefficiency correspond to fluctuations in unpredictability: during efficient (sideways) phases unpredictability is maximal, whereas inefficiencies introduce positive or negative autocorrelation in returns and hence predictability: positive autocorrelation generates momentum regimes and lower volatility than the efficient benchmark, while negative autocorrelation produces reversal regimes and higher volatility. In both cases, unpredictability decreases with respect to the baseline efficient scenario, where it is the unpredictability and not the volatility to be maximum.\\

Introducing the Multifractional Processes with Random Exponents (MPRE), a very general class of stochastic processes that can model both the efficient and inefficient phases in financial markets, we obtain two main results:  
\begin{enumerate}[leftmargin=*]
    \item we deduce an analytical relationship between regularity (predictability) and volatility, yielding a theoretically grounded “fair 
    volatility’’ benchmark;
    \item we provide extensive empirical evidence from fourteen global equity indices supporting the model and enabling a quantitative assessment of deviations from market efficiency, distinguishing momentum from mean-reversion phases.
\end{enumerate}

The remainder of the paper is structured as follows. Section~\ref{sec:Model} reviews the pointwise Hölder exponent and its relation with semimartingales.  
Section~\ref{sec:MPRE} introduces the MPRE and presents the main result linking the Hurst-Hölder exponent to the local volatility.
Section~\ref{sec:RiskVol} provides the financial interpretation of the Hurst–Hölder regularity. Section~\ref{sec:Hestimation} is devoted to some key properties of established Hurst–Hölder estimators and to a Monte Carlo analysis of a stressed estimation. Section~\ref{sec:Application} presents an extensive empirical analysis on 14 stock indices supporting the theoretical results. Section~\ref{sec:Conclusion} concludes.

\section{Pointwise H\"older exponent and semimartingales}
\label{sec:Model}
This section introduces the pointwise Hölder exponent as a preliminary tool essential for understanding the developments that follow. 
%\subsection{Pointwise H\"older exponent and semimartingales}
\begin{defn}[Pointwise Hölder exponent of a stochastic process]
Let $Y=\{Y_t : t\in I\}$ be a real-valued stochastic process defined on an interval $I\subset\mathbb{R}$, and assume that $Y$ is almost surely continuous at $t_0\in I$.
For $\alpha>0$, we say that $Y$ is Hölder continuous of order $\alpha$ at $t_0$ if, with probability one,
\begin{equation*}
    \limsup_{s\to t_0} 
  \frac{|Y_s - Y_{t_0}|}{|s - t_0|^\alpha} < \infty.
\end{equation*}
The pointwise Hölder exponent of $Y$ at $t_0$ is the (possibly random) variable
\begin{equation*}
    \alpha_Y(t_0)
  := \sup\Big\{
      \alpha>0:\ 
      \limsup_{s\to t_0}
      \frac{|Y_s - Y_{t_0}|}{|s - t_0|^{\alpha}}
      < \infty 
      \ \text{a.s.}
    \Big\}.
\end{equation*}
Equivalently, $\alpha_Y(t_0)\geq \alpha$ almost surely if and only if the sample-path increments 
satisfy
\begin{equation*}
   |Y_s - Y_{t_0}| = O\!\left(|s-t_0|^\alpha\right)
   \qquad\text{as } s\to t_0, 
\end{equation*}
and $\alpha_Y(t_0)<\alpha$ almost surely if and only if
\begin{equation*}
    \limsup_{s\to t_0}
   \frac{|Y_s - Y_{t_0}|}{|s-t_0|^\alpha}
   = \infty.
\end{equation*}
\end{defn}

Notice that for all diffusions with nondegenerate, progressively measurable and locally bounded $\mu_t$ and $\sigma_t \neq 0$ (bounded away from zero at any time $t_0$), the Hölder exponent %of the Itô SDE
%\begin{equation*}
%    dX_t = \mu_t\,dt\,+\,\sigma_t\,dW_t
%\end{equation*}
is $\frac{1}{2}$ almost surely. This result stems from the following argument. Consider the continuous semimartingale written as
\begin{equation*}
     X_t = \int_0^t \mu_s\,ds + \int_0^t \sigma_s\,dW_s,\qquad t\in[0,T],
\end{equation*}
and denote by $A_t := \int_0^t \mu_s\,ds$ the continuous finite-variation process and by $M_t := \int_0^t \sigma_s\,dW_s$ the continuous local martingale.  
The Hölder regularity of $X$ at a given time $t_0$, denoted by $\alpha_X(t_0)$, follows from that of these two components.\\ 
Since $\mu$ is locally bounded near $t_0$ there exists
$K>0$ and $\delta>0$ such that for all $s\in[t_0-\delta,t_0+\delta]$,
$|\mu_s|\leq K$ almost surely.\\
For $|t-t_0|\leq\delta$, we have
\begin{equation*}
    |A_t - A_{t_0}|
  = \Big|\int_{t_0}^t \mu_s\,ds\Big|
  \le \int_{t_0}^t |\mu_s|\,ds
  \le K\,|t-t_0|.
\end{equation*}
Thus, for any $\alpha\leq 1$,
\begin{equation*}
    \frac{|A_t-A_{t_0}|}{|t-t_0|^\alpha} \le K |t-t_0|^{1-\alpha},
\end{equation*}
which is bounded in a neighbourhood of $t_0$ whenever $\alpha \leq 1$.
Therefore,
\begin{equation*}
    \alpha_A(t_0) \geq 1.
\end{equation*}

On the other hand, by the Dambis--Dubins--Schwarz theorem there exists a (possibly enlarged) probability space and a standard Brownian motion $B$ such that
\begin{equation*}
    M_t = B_{\langle M\rangle_t}, \qquad 0\le t\le T,
\end{equation*}
 i.e. $M$ is a time-changed Brownian motion (here as usual  $\langle M \rangle_t$ denotes the quadratic variation of $M$). Classical results based on the law of the iterated logarithm and the exact modulus of continuity imply that the Hölder exponent of standard Brownian motion at any time $t_0$ is $\frac{1}{2}$ almost surely. In particular, for the time-changed martingale $M_t$, we have that:

\begin{itemize}[leftmargin=*]
    \item if the quadratic variation increases at $t_0$, i.e. 
    $\frac{d}{dt}\langle M\rangle_t\big|_{t=t_0} > 0$, then
    \begin{equation*}
        \alpha_M(t_0) = \frac{1}{2};
    \end{equation*}
    \item if $\langle M\rangle_t$ is constant on a neighborhood of $t_0$, i.e. $M_t = M_{t_0}$ for all $t$ sufficiently close to $t_0$, then
    \begin{equation*}
        \alpha_M(t_0) = +\infty.
    \end{equation*}
\end{itemize}
Thus, the Hölder exponent equals $1/2$ precisely at points of genuine stochastic activity---that is, points where the quadratic variation increases.\\

Since for continuous functions $Y,Z$ one has
\begin{equation*}
    \alpha_{Y+Z}(t_0) = \min\{\alpha_Y(t_0),\, \alpha_Z(t_0)\},
\end{equation*}
the rougher of the two components determines the local behaviour of the sum. Therefore,
\begin{equation*}
    \alpha_X(t_0) = \min\{\alpha_A(t_0),\alpha_M(t_0)\} = \frac{1}{2} \qquad\text{a.s.},
\end{equation*}
independently of the value of $\sigma_{t_0}$.
%Thus, excluding the degenerate cases in which the the quadratic variation is flat, the pointwise Hölder exponent of every continuous local martingale is $\frac{1}{2}$ (at points where the quadratic variation increases).
%As a consequence, efficient markets may display either “low’’ volatility—suggesting smooth price adjustment to new information—or “high’’ volatility—indicating rapid incorporation of complex information—yet theory provides no criterion for distinguishing these states in terms of what is “normal.’’ This lack of a reference level limits the usefulness of volatility in regulatory and risk management contexts: without a benchmark separating ordinary market conditions from excessive risk-taking, policy interventions become difficult to calibrate. These challenges are compounded by the distinction between realized and implied volatility and by the complex term structures associated with volatility expectations \citep{Bucci2020}. Expected volatility may differ substantially from realized volatility, introducing further uncertainty \citep{christensen1998relation}.

\section{Multifractional Process with Random Exponent}\label{sec:MPRE}
%A different strategy to model stochastic volatility replaces system \eqref{eq:SVgeneral} by a \textit{Multifractional Process with Random Exponent} (MPRE). 
Originated by the need to overcome the limitations of a constant Hurst exponent\footnote{Notable extensions include the multifractional Brownian motion (mBm) \cite{PL1995,BJR1997} and its generalized variant (GmBm) \cite{AL2000}, bifractional Brownian motion \cite{HV2003}, mixed fBm \cite{Che2001}, fractional Riesz-Bessel motion \cite{Anh1999}, the Multi-fractional Generalized Cauchy Process \cite{Li2020}, and Multifractional Processes with Random Exponents (MPRE) \cite{AT2005} (see \cite{Lim2015} for a comprehensive review). For other interesting directions, see also \citep{Eliazar2013, Eliazar2024}.}, 
%The assumption of a constant Hurst exponent is often empirically untenable, particularly for processes characterized by structural breaks, regime shifts, or heteroskedastic dynamics—hallmark features of stochastic volatility models. To overcome this limitation inherent in fBm, the framework has been extended through several generalizations\footnote{Notable extensions include the multifractional Brownian motion (mBm) \cite{PL1995,BJR1997} and its generalized variant (GmBm) \cite{AL2000}, bifractional Brownian motion \cite{HV2003}, mixed fBm \cite{Che2001}, fractional Riesz-Bessel motion \cite{Anh1999}, the Multi-fractional Generalized Cauchy Process \cite{Li2020}, and Multifractional Processes with Random Exponents (MPRE) \cite{AT2005} (see \cite{Lim2015} for a comprehensive review). For other interesting directions, see also \citep{Eliazar2013, Eliazar2024}.}.
%Among these, one of the most flexible and encompassing frameworks is provided by 
the MPRE was firstly introduced in the seminal work of
\citep{AT2005, AJT2007}, who constructed it via random wavelet series and proved that—under mild conditions—the local roughness (or H\"{o}lder regularity) of the sample paths at a point $t$ is directly governed by the value of the functional parameter $H_t$, thus justifying its designation as the Hurst-H\"{o}lder exponent.

A significant constraint of this original MPRE construction, however, is its incompatibility with the standard Itô calculus framework. To remedy this drawback, a subsequent MPRE formulation was introduced by \citep{AEH2018} and further generalized in \citep{Lobodaetal2021}. This latter class of processes will serve as the foundational model for the subsequent analysis. Although the framework is very general, % given the filtered probability space $(\Omega, \pazccal{F}, (\pazccal{F}_s)_{s\in \mathbb{R}}, \mathbb{P})$,
we will refer to the following specification of MPRE:%, which generalizes equation \eqref{eq:fbm}:

\begin{equation} \label{eq:MPRE1}
X_t = \int_{-\infty}^t \nu_s \left[(t-s)_+^{H_s-1/2}-(-s)_+^{H_s-1/2}\right] dW_s,
\end{equation}
where $\nu_s>0$ is a $\pazccal{F}_s$-adapted, time-varying scale coefficient, almost surely continuous and bounded; $H_s\in(0,1)$ is a random variable or even an $\pazccal{F}_s$-adapted stochastic process and $(\cdot)_+ = \max(0,\cdot)$.

\begin{remark}[Generalization of fBm] When $\nu_t=\nu$ and $H_t=H$, the MPRE trivially reduces to the fractional Brownian Motion (fBm) \citep{Kolmogorov1940,MV1968}%, denoted $W^H_t$, is a continuous-time Gaussian process starting at zero with stationary increments, parameterized by a Hurst exponent \(H \in (0,1)\). Its integral representation is
\begin{align}\label{eq:fbm}
    W^H_t 
    &= \frac{1}{\Gamma\!\left(H+\tfrac{1}{2}\right)} 
       \int_{-\infty}^t \left[(t-s)^{H-1/2}_+ - (-s)^{H-1/2}_+\right]\, dW_s,
\end{align}
 with variance of increments given by
\begin{equation} \label{eq:varfBm}
    \mathbb{E}\!\left[(W^H_{t+T}-W^H_t)^2\right] = V_H T^{2H}.
\end{equation}
Here $V_H$ is the variance of unit-lag increments, explicitly \citep{Reed1995, DU1999}
\begin{eqnarray*}
    V_H %&=& \frac{1}{\Gamma\!\left(H+\tfrac{1}{2}\right)^2}\left\{\frac{1}{2H} + \int_1^\infty \!\left[u^{H-1/2}-(u-1)^{H-1/2}\right]^2 du \right\},\\
    %&=& \frac{\Gamma(1-2H)}{\pi H} \cos(\pi H)\\
    %&=& \frac{\Gamma(H)\Gamma(1-H)}{\pi \Gamma(1+2H)} \\
    %&=& \frac{\Gamma(2-2H)\cos(\pi H)}{\pi H(1-2H)} \\
    &=& \frac{1}{2H\sin(\pi H)\Gamma(2H)}=:\frac{A(H)}{\Gamma(H+1/2)^2}.
\end{eqnarray*}
%The above equalities summarize the different expressions that can be found in literature for $V_H$ and are explictly proved in \citep{Bianchietal2025}. In particular, 
In the last equality, $A(H)$ is the quantity considered for the covariance function by \citep{Lobodaetal2021} (equation (28)). This relation will be useful in Proposition \ref{prop:1} to characterize the local behavior of MPRE.
%The covariance function reads as
%\[
%    \mathbb{E}[W^H_t W^H_s] = \frac{V_H}{2}\left(|t|^{2H} + |s|^{2H} - |t-s|^{2H}\right), \quad t,s \geq 0.
%\]
%The parameter $H$ fully characterizes dependence structure and path regularity. The case $H=1/2$ corresponds to standard Brownian motion with independent increments. When $H>1/2$, increments are positively correlated, generating persistence and long memory, while $H<1/2$ induces negative correlation of increments, associated with anti-persistence and short memory. 
% In the particular case where all the random variables $H_s$ are equal to the same deterministic constant $H$, $X_t$ reduces to $W^H_t$.
\end{remark}

\begin{remark}[H\"older regularity]
    Loboda et al. \citep{Lobodaetal2021} show (see Theorem 3.4 of Section 3 of their work) that $P(\alpha_X(t) = H_t)=1$. i.e. at any time $t$ the pointwise Hölder exponent of MPRE is almost surely $H_t$ if $H_t$ is continuous, irrespective of the regularity of $H_t$. For the case that $H_t$ is discontinuous, similar results which are only slightly weaker are obtained in the same reference. 
\end{remark}

\begin{remark}[Covariance]
   Assuming that $H_t$ and $\nu_t$ are stationary and $\pazccal{F}_s$-adapted, \citep{Lobodaetal2021} prove that
    \begin{equation} \label{eq:ACFMPRE}
        \mathbb{E}(X_tX_s) = \mathbb{E}\left[\nu_0^2\frac{A_{H_0}}{2}\left(|t|^{2H_0}+|s|^{2H_0}-|t-s|^{2H_0}\right)\right]
    \end{equation}
    provided that the latter value is finite for all $t,s\in\mathbb{R}$. 
\end{remark}

 \begin{figure}[H]
        \centering
\includegraphics[width=\linewidth,height=0.3\textheight,keepaspectratio]{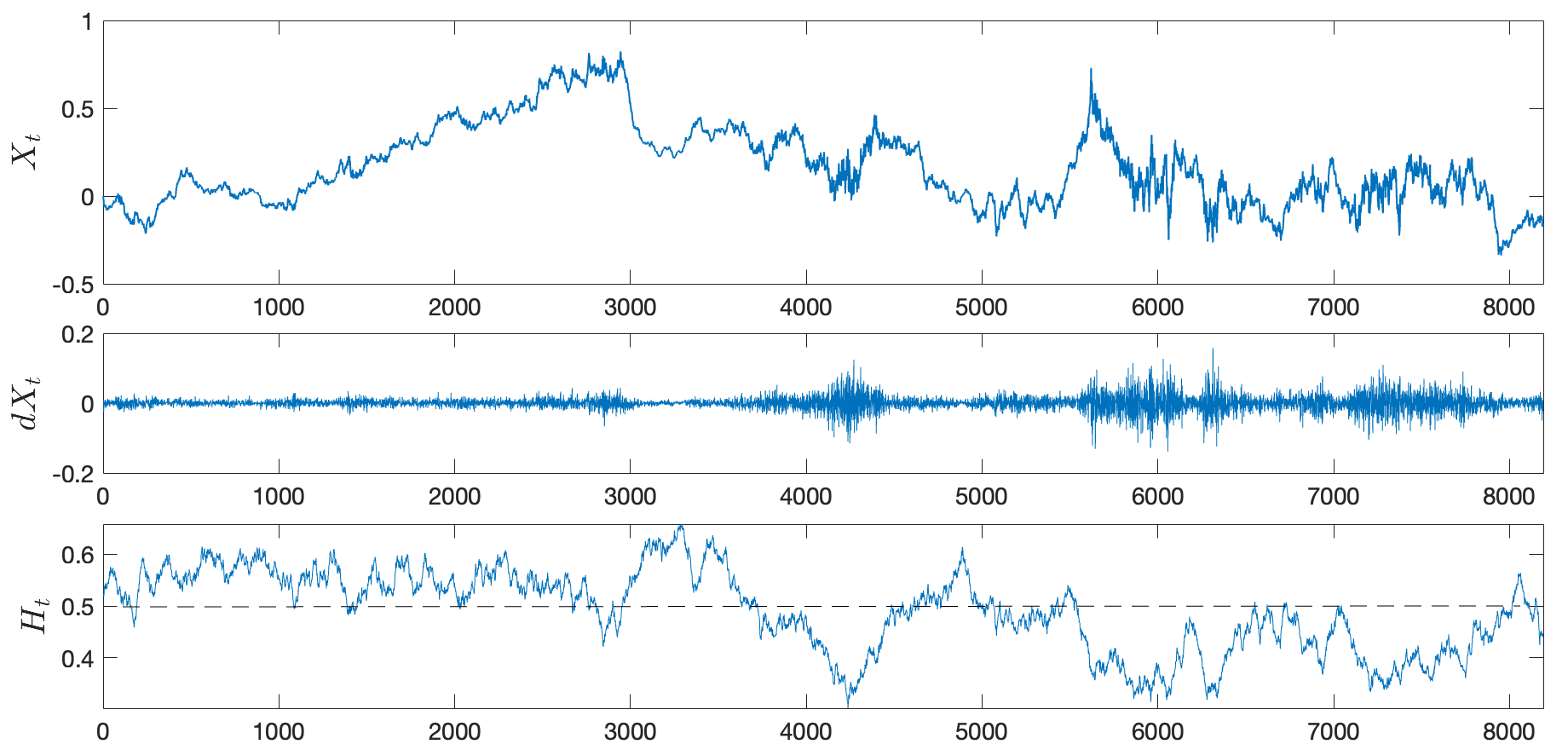}
\captionsetup{margin=1.9cm}
       \caption{Example of a path of MPRE with $\nu_t=1$ and $H_t$ generated by the Ornstein-Uhlenbeck process $dH_t=\kappa(\theta-H_t)\,dt+\sigma\,dW_t$ with $\kappa = 0.2, \theta = 1/2$ and $\sigma=0.05$.} \label{fig:exampleMPRE}
    \end{figure}

Although it is not yet widely used in financial modeling, the MPRE provides a very general, flexible and parsimonious model of the asset price dynamics (see e.g. \citep{CLV2014,Peng2018,Pengetal2022}, for theoretical arguments; and \citep{CV2003,Garcin2017,AmmyGarcin2023,Chong2025} for extensive empirical research). Furthermore, the randomization of the regularity parameter is in line with similar modeling choices already introduced to fit the skewness of the volatility surface \citep{Carr2007,Grzelach2026}.

Figure \ref{fig:exampleMPRE} clearly shows why: using only function $H_t$, which has an immediate, intuitive and robust financial meaning (see Section \ref{sec:RiskVol} for a detailed discussion), the model is able to reproduce virtually every stylized fact known about asset price dynamics. Its main appeal lies in the ability to accommodate both efficient market regimes, associated with a Hurst-Hölder exponent $H_t=1/2$, and periods of disequilibrium, where $H_t>1/2$ reflects momentum and $H_t<1/2$ reflects mean reversion. The reason why this occur is simple: the kernel $(t-s)_+^{H_s-1/2}-(-s)_+^{H_s-1/2}$ controls how much past noise is recycled into the present. Since $X_t$ behaves locally as a fBm with Hurst exponent $H_t$, when $H_t>1/2$, the local Volterra kernel assigns relatively greater weight to past increments, yielding smoother trajectories and inducing positive short-horizon dependence: upward or downward movements tend to be followed by further movements in the same direction, consistent with local momentum. Conversely, when $H_t<1/2$, the kernel is locally rough and overemphasizes recent shocks relative to older ones, generating negative short-horizon dependence: increments tend to overshoot and subsequently revert, producing local antipersistence or mean reversion.\\

Assuming asset prices evolve according to an MPRE, a natural question concerns the link between the pointwise variance of the process and the time-varying Hurst function $H_t$. Establishing this connection is central to our analysis: it ensures that the volatility measure $\sigma_t$ remains path-dependent and sensitive to local dynamics. In this way, a volatility measure based on $H_t$ retains essential information on time-varying memory properties, offering a richer characterization of latent risk than conventional specifications. To address the above question, we prove the following%, which generalizes equation (15) in \citep{Bianchietal2025} for a lag $h\neq 1$.

\begin{prop}[Standard deviation of MPRE increments] \label{prop:1}
Let:
\begin{itemize}
\item[\textbf{H1}] $W$ in \eqref{eq:MPRE1} be a standard Brownian motion independent of $(H,\nu)$;
\item[\textbf{H2}] $H$ and $\nu$ be almost surely continuous at $t$ with values in a compact subset $(\underline H,\overline H)\subset(0,1)$ and $[\underline\nu,\overline\nu]\subset(0,\infty)$;
\item[\textbf{H3}] $\int_{-\infty}^{t}\nu^2_s\big[(t-s)_{+}^{H_s-\frac{1}{2}}-(-s)_{+}^{H_s-\frac{1}{2}}\big]^2ds<\infty$ a.s. (this holds under \textbf{H1}–\textbf{H2}),
\end{itemize}
then the increments of the MPRE in equation \eqref{eq:MPRE1}
%\begin{equation*}
%    X(t)=\int_{-\infty}^{t}\nu(s)\Big[(t-s)_{+}^{H(s)-\frac{1}{2}}-(-s)_{+}^{H(s)-\frac{1}{2}}\Big]\,dW(s),
%\end{equation*}
%with a random Hurst function $H(s)\in(0,1)$ and a volatility factor $\nu(s)>0$ 
have (conditional) standard deviation given by
\begin{equation} \label{eq:IncrVar}
\mathrm{sd}\!\left(X_{t+h}-X_t\,\big|\,\pazccal{F}_t^{H,\nu}\right)\sim
|h|^{H_t}\,\nu_t\,\sqrt{A_{H_t}}\,,\qquad h\to0,    
\end{equation}
where $\pazccal{F}_t^{H,\nu}:=\sigma\!\big(\{(H_s,\nu_s):s\le t\}\big)$ is the $\sigma$–field generated by $(H,\nu)$ up to time $t$ and
\begin{equation*}
A_H=\frac{\Gamma\!\left(H+\tfrac12\right)^2}{2H\,\sin(\pi H)\,\Gamma(2H)}.    
\end{equation*}
%Here $\pazccal{F}_t^{H,\nu}:=\sigma\!\big(\{(H_s,\nu_s):s\le t\}\big)$ is the $\sigma$–field generated by $(H,\nu)$ up to time $t$.    
\end{prop}
\vspace{.1cm}
\begin{proof}
We prove the result for $h>0$; the case $h<0$ then follows by symmetry, yielding the factor $|h|^{H_t}$. We split the proof into the six following steps.
%\end{flushleft}

\bigskip

\noindent\textbf{Step 1: Increment decomposition.}
For $h>0$,
\begin{align*}
X_{t+h}-X_t
&=\int_{-\infty}^{t+h}\!\!\nu_s\Big[(t{+}h-s)_{+}^{H_s-\frac12}-(-s)_{+}^{H_s-\frac12}\Big]\,dW_s
 -\int_{-\infty}^{t}\!\!\nu_s\Big[(t-s)_{+}^{H_s-\frac12}-(-s)_{+}^{H_s-\frac12}\Big]\,dW_s\\
&=\int_{-\infty}^{t}\nu_s\Big[(t{+}h{-}s)^{H_s-\frac12}-(t{-}s)^{H_s-\frac12}\Big]\,dW_s
  \;+\;\int_{t}^{t+h}\nu_s\,(t{+}h{-}s)^{H_s-\frac12}\,dW_s,
\end{align*}
since the $(-s)_+^{H_s-\frac12}$ terms cancel for $s<0$, and are zero for $s>0$.

\bigskip

\noindent\textbf{Step 2: Conditional Itô isometry.}
By \textbf{H1}, conditioning on $\pazccal{F}_t^{H,\nu}$ makes the integrands deterministic while preserving the Gaussianity/zero mean of the Itô integrals. Hence, \textbf{H3} ensuring the convergence of integral, one has
\begin{align}
\operatorname{Var}\!\left(X_{t+h}-X_t\,\big|\,\pazccal{F}_t^{H,\nu}\right)
&=\int_{-\infty}^{t}\!\nu_s^2\Big[(t{+}h{-}s)^{H_s-\frac12}-(t{-}s)^{H_s-\frac12}\Big]^2\,ds
 +\int_{t}^{t+h}\!\nu_s^2\,(t{+}h{-}s)^{2H_s-1}\,ds. \label{eq:condvar}
\end{align}

\medskip

\noindent\textbf{Step 3: Scaling change of variables.}
Set, for $s\le t$, $u=(t-s)/h\in(0,\infty)$, i.e. $s=t-hu$, $ds=-h\,du$; and, for $s\in(t,t+h]$, $v=(t+h-s)/h\in(0,1)$, $ds=-h\,dv$. Then \eqref{eq:condvar} becomes
\begin{align}
\operatorname{Var}\!\left(X_{t+h}-X_t\,\big|\,\pazccal{F}_t^{H,\nu}\right)
&=h^{2H_t}\Bigg[
\int_{0}^{\infty}\!\!\nu_{t-hu}^2\,h^{2(H_{t-hu}-H_t)}
\Big[(u{+}1)^{H_{t-hu}-\frac12}-u^{H_{t-hu}-\frac12}\Big]^2\,du \nonumber\\
&\qquad\qquad\quad+\int_{0}^{1}\!\!\nu_{t+h(1-v)}^2\,h^{2(H_{t+h(1-v)}-H_t)}\,v^{2H_{t+h(1-v)}-1}\,dv\Bigg].
\label{eq:scaled}
\end{align}

\medskip

\noindent\textbf{Step 4: Local freezing (continuity) and dominated convergence.}
By \textbf{H2}, as $h\downarrow0$ we have
\[
\nu_{t-hu}\to\nu_t,\qquad H_{t-hu}\to H_t\qquad\text{and}\qquad
\nu_{t-h(1-v)}\to\nu_t,\; H_{t-h(1-v)}\to H_t,
\]
uniformly over $u$ in compact subsets of $(0,\infty)$ and $v$ in $[0,1]$.\\
Moreover, for all sufficiently small $h$,
\[
\underline H< H_{t-hu},H_{t-h(1-v)}<\overline H,
\qquad
\underline\nu\le \nu_{t-hu},\nu_{t-h(1-v)}\le \overline\nu.
\]
Hence the two integrands in \eqref{eq:scaled} are dominated respectively by
\[
C\Big[(u{+}1)^{\overline H-\frac12}-u^{\underline H-\frac12}\Big]^2
\quad\text{and}\quad
C\,v^{2\underline H-1},
\]
for a constant $C>0$ independent on $h$; both are integrable on $(0,\infty)$ and $(0,1)$ because, as $u\to\infty$, the difference behaves like $u^{\overline H-\frac32}$ and thus its square is $u^{2\overline H-3}$ (integrable since $\overline H<1$), while as $u\downarrow0$ the term $u^{\underline H-\frac12}$ yields square $u^{2\underline H-1}$ (integrable as $\underline H>0$), and $v^{2\underline H-1}$ is integrable on $(0,1)$.

Passing to the limit $h\downarrow0$ inside the brackets in \eqref{eq:scaled} by dominated convergence, we obtain
\begin{align}
\operatorname{Var}\!\left(X_{t+h}-X_t\,\big|\,\pazccal{F}_t^{H,\nu}\right)
&=h^{2H_t}\nu_t^{2}\Bigg[
\underbrace{\int_{0}^{\infty}\!\Big((u{+}1)^{H_t-\frac12}-u^{H_t-\frac12}\Big)^{2}du}_{=:J(H_t)}
\;+\;\underbrace{\int_{0}^{1}\!v^{2H_t-1}dv}_{=\,\frac{1}{2H_t}}
\Bigg]\;+\;o\!\left(h^{2H_t}\right) \nonumber\\
&=h^{2H_t}\nu_t^{2}\Big(J(H_t)+\tfrac{1}{2H_t}\Big)+o\!\left(h^{2H_t}\right).
\label{eq:var-asymp}
\end{align}

Thus it remains to compute the constant
\begin{equation*}
    J(H):=\int_{0}^{\infty}\!\Big((u{+}1)^{H-\frac12}-u^{H-\frac12}\Big)^{2}du
\qquad H\in(0,1).
\end{equation*}

\medskip

\noindent\textbf{Step 5: Evaluation of $J(H)$.}
Notice that $J(H)$ corresponds to the contribution from $s\le t$; the second integral in \eqref{eq:var-asymp} supplies the contribution from $s\in(t,t+h]$. A direct way to evaluate $J(H)$ is via the Fourier transform of the kernel
\begin{equation*}
    k_H(u):=u_{+}^{\,H-\frac12},\qquad u\in\mathbb R,
\end{equation*}
whose (tempered–distribution) Fourier transform is the classical formula (for $0<H<1$)
\begin{equation*}
    \widehat{k_H}(\omega)=\int_{0}^{\infty}u^{H-\frac12}e^{-i\omega u}\,du
= e^{-i\,\mathrm{sgn}(\omega)\,\frac{\pi}{2}\left(H+\frac12\right)}\,
\Gamma\!\Big(H+\tfrac12\Big)\,|\omega|^{-H-\frac12}.
\end{equation*}
Let $D_hk_H(u):=k_H(u+h)-k_H(u)$. By Parseval–Plancherel identity,
\begin{equation*}
    \int_{\mathbb R}\big(D_hk_H(u)\big)^2\,du
=\frac{1}{2\pi}\int_{\mathbb R}\big|(e^{i\omega h}-1)\,\widehat{k_H}(\omega)\big|^2\,d\omega
=\frac{\Gamma\!\left(H+\tfrac12\right)^2}{2\pi}\int_{\mathbb R}|e^{i\omega h}-1|^2\,|\omega|^{-2H-1}\,d\omega.
\end{equation*}
Since $|e^{i\omega h}-1|^2=2(1-\cos(\omega h))$ and the integrand is even,
\begin{align}
\int_{\mathbb R}\big(D_hk_H(u)\big)^2\,du
&=\frac{2\Gamma\!\left(H+\tfrac12\right)^2}{\pi}\int_{0}^{\infty}\!(1-\cos(\omega h))\,\omega^{-2H-1}\,d\omega \nonumber\\
&=\frac{2\Gamma\!\left(H+\tfrac12\right)^2}{\pi}\,h^{2H}
\underbrace{\int_{0}^{\infty}\!(1-\cos x)\,x^{-2H-1}\,dx}_{=:I(H)}.
\label{eq:Plancherel}
\end{align}

\medskip

\noindent\textbf{Lemma (cosine integral).}
For $H\in(0,1)$,
\begin{equation*}
    I(H):=\int_{0}^{\infty}(1-\cos x)\,x^{-2H-1}\,dx
=\frac{\pi}{2\,\Gamma(2H+1)\,\sin(\pi H)}.
\end{equation*}

\noindent \textbf{Proof.}
Consider the classical identity (valid for $0<\alpha<2$)
\[
\int_{0}^{\infty}(1-\cos x)\,x^{-1-\alpha}\,dx
=-\Gamma(-\alpha)\,\cos\!\Big(\frac{\pi\alpha}{2}\Big),
\]
which follows by analytic continuation from
$\int_{0}^{\infty}x^{\beta-1}\cos x\,dx=\Gamma(\beta)\cos(\pi\beta/2)$ for $\beta\in(0,1)$.
Setting $\alpha=2H$ yields
\[
I(H)=-\Gamma(-2H)\,\cos(\pi H).
\]
Using $\Gamma(2H+1)=2H\,\Gamma(2H)$ and the reflection identity
$\Gamma(-2H)\Gamma(2H+1)=-\pi/\sin(2\pi H)$, we get
\begin{equation*}
    -\Gamma(-2H)\,\cos(\pi H)=\frac{\pi}{\Gamma(2H+1)}\cdot\frac{\cos(\pi H)}{\sin(2\pi H)}
=\frac{\pi}{2\,\Gamma(2H+1)\,\sin(\pi H)}.
\qedhere
\end{equation*}

\medskip

Substituting $I(H)$ into \eqref{eq:Plancherel} gives
\[
\int_{\mathbb R}\big(D_hk_H(u)\big)^2\,du
=\frac{2\Gamma\!\left(H+\tfrac12\right)^2}{\pi}\,h^{2H}\cdot
\frac{\pi}{2\,\Gamma(2H+1)\,\sin(\pi H)}
=\frac{\Gamma\!\left(H+\tfrac12\right)^2}{\,\Gamma(2H+1)\,\sin(\pi H)}\,h^{2H}.
\]
On the other hand, a direct decomposition of the left–hand side over the regions $u>0$, $-1<u\le0$, $u\le-1$ yields
\[
\int_{\mathbb R}\big(D_hk_H(u)\big)^2\,du
=h^{2H}\left(\int_{0}^{\infty}\!\Big((u{+}1)^{H-\frac12}-u^{H-\frac12}\Big)^2du
+\int_{0}^{1}\!v^{2H-1}\,dv\right)
=h^{2H}\Big(J(H)+\tfrac{1}{2H}\Big).
\]
Equating the two expressions and using $\Gamma(2H+1)=2H\,\Gamma(2H)$ we obtain
\[
J(H)+\frac{1}{2H}
=\frac{\Gamma\!\left(H+\tfrac12\right)^2}{\Gamma(2H+1)\,\sin(\pi H)}
=\frac{\Gamma\!\left(H+\tfrac12\right)^2}{2H\,\sin(\pi H)\,\Gamma(2H)}
=:A_H.
\]
This is the desired constant.

\medskip

\noindent\textbf{Step 6: Conclusion (variance and standard deviation).}
Combining \eqref{eq:var-asymp} with the evaluation of $J(H)$ we have, as $h\downarrow0$,
\begin{equation*}
\operatorname{Var}\!\left(X_{t+h}-X_t\,\big|\,\pazccal{F}_t^{H,\nu}\right)
= h^{2H_t}\,\nu_t^{2}\,A_{H_t}+\;o\!\left(h^{2H_t}\right).
\end{equation*}
Taking square roots (and using that $A_{H_t}>0$ and $h^{2H_t}\to0$ so that the ratio of the square roots equals the square root of the ratio), we conclude
\begin{equation*}
    \mathrm{sd}\!\left(X_{t+h}-X_t\,\big|\,\pazccal{F}_t^{H,\nu}\right)
\sim |h|^{H_t}\,\nu_t\,\sqrt{A_{H_t}}\,,\qquad h\to0,
\end{equation*}
which proves the claim.
\begin{flushright}
\qedsymbol{}
\end{flushright}
\end{proof}

\noindent \textbf{Remarks.}
\begin{itemize}[leftmargin=*]
%\item The proof crucially used \textbf{H1} to apply the conditional Itô isometry with respect to $\pazccal{F}_t^{H,\nu}$. The continuity assumption \textbf{H2} yields the ``local freezing'' $\nu_{t\pm o(1)}\to\nu_t$ and $H_{t\pm o(1)}\to H_t$, which justifies dominated convergence in Step~4.
\item The constant $A_H$ coincides with the variance of the unit increment of the (non–normalised) fractional Brownian motion associated with the Mandelbrot–Van Ness kernel used here; equivalently, if one rescales the kernel by the factor $c_H:=\Gamma\!\left(H+\tfrac12\right)^{-1}\sqrt{2H\sin(\pi H)\Gamma(2H)}$, then $\operatorname{Var}(W^H_{t+h}-W^H_t)=|h|^{2H}$.
\item Standard deviation \eqref{eq:IncrVar} is consistent with the autocovariance function \eqref{eq:ACFMPRE}, from which it directly follows the autocovariance function $\gamma(h)$ of the increment  process
    \begin{equation*}
    \gamma(h)=\mathbb{E}\left[\nu_0^2\frac{A_{H_0}}{2}\left(|h+1|^{2H_0}-2|h|^{2H_0}+|h-1|^{2H_0}\right)\right]
    \end{equation*} 
    In particular, equation \eqref{eq:IncrVar} is a small-$h$ asymptotic for the general MPRE; when $\nu_t$ and $H_t$ are stationary, the equality holds exacly for all $h$ as $\mathbb{E}[(X_{t+h}-X_t)^2]=\mathbb{E}[|h|^{2H_0}\nu_0^2A_{H_0}]$, as stated in the proof provided by \citep{Lobodaetal2021}.
\end{itemize}

\begin{figure}%[hbp]
\centering
% -------- LEFT: FIGURE (0.60\textwidth) --------
\begin{minipage}[H]{0.50\textwidth}
  \vspace{0pt}%
  \centering
  % Constrain caption to the minipage width (prevents overflow)
  %\captionsetup{width=\linewidth}
  \captionsetup{margin=.5cm}
        \includegraphics[width=\linewidth,height=0.3\textheight,keepaspectratio]{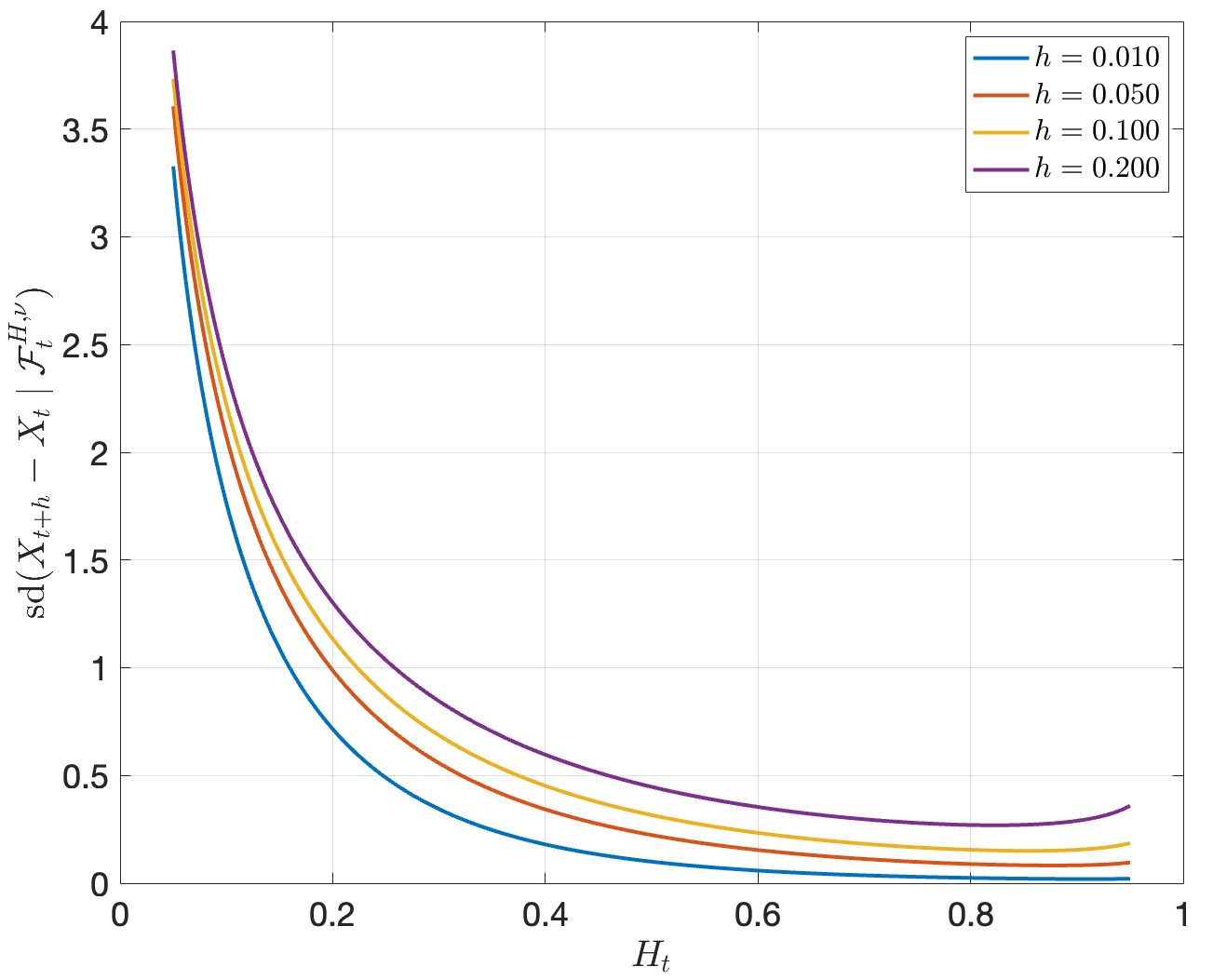}
       \caption{Asymptotic conditional standard deviation vs. $H_t$ for different values of increments $h$} \label{fig:AsCondSD}
\end{minipage}%
\hfill
% -------- RIGHT: TABLE (0.37\textwidth) --------
\begin{minipage}[H]{0.50\textwidth}
  \vspace{-50pt}%
  \centering
  \captionsetup{margin=1.4cm}
  \small
  \captionof{table}{Numerical minima of 
$\mathrm{sd}(H,h) \approx |h|^H \sqrt{A_H}$ (with $\nu_t=1$) 
over $H\in(0,1)$, for several values of $h$.}
  \label{tab:minima}
  \begin{tabular}{ccccc}
\toprule
$h$ & & $H^*(h)$ & &$\mathrm{sd}_{\min}(h)$ \\
\hline
$0.200$ & & $0.823$ & & $0.2690$ \\
$0.100$ & & $0.855$ & & $0.1503$ \\
$0.050$ & & $0.878$ & & $0.0824$ \\
$0.020$ & & $0.900$ & & $0.0365$ \\
$0.010$ & & $0.912$ & & $0.0195$ \\
$0.005$ & & $0.921$ & & $0.0103$ \\
$0.001$ & & $0.937$ & & $0.0023$ \\
\bottomrule
\end{tabular}
\end{minipage}
\end{figure}

For each fixed $|h|<1$, the function $H \mapsto |h|^{H}\sqrt{A_H}$ can have an interior minimum $H^*(h)$, because the non-monotonic behavior of $A_H$ in $H$ competes with the monotone decay of $|h|^H$. Only in the asymptotic regime $|h| \to 0$ does the decay in $|h|^H$ dominate enough that the minimum is effectively pushed towards higher values of $H$ (see Figure \ref{fig:AsCondSD} and Table \ref{tab:minima}). The fact that the standard deviation \eqref{eq:IncrVar} is injective up to $H^*(h)$ ensures in the application of Section \ref{sec:Hestimation} that the estimated pointwise regularity can be converted into the corresponding volatility, since the estimated maximum $H_t$ is in any case less than $H^*(h)$. This makes it possible to identify a fair volatility, which -- consistent with the semimartingale assumption -- varies from series to series.

\section{$H_t$ as a risk indicator bridging rational and behavioral finance} \label{sec:RiskVol}
In the previous Sections we have discussed how the Hurst-Hölder exponent $H_t$ measures pointwise regularity and quantifies deviations from the semimartingale framework (see e.g. \citep{LeRoy1973,Lucas1978} for the derivation and the connection with the economic definition of efficiency). Its critical value, $H_t = 1/2$, corresponds to martingale dynamics, making it a natural foundation for modeling volatility in non-semimartingale settings. This approach offers distinct advantages over conventional volatility measures:

\paragraph*{a) Sensitivity to Autocorrelation Structure}
Unlike volatility, which is a scale-dependent measure of dispersion, the Hurst-H\"{o}lder exponent is scale-invariant and directly responsive to the autocorrelation structure of returns. Processes with differing temporal dependencies can exhibit identical volatility yet possess distinct $H_t$ values, as autocorrelation intrinsically governs the regularity of sample paths. Under the MPRE specification, Proposition 1 implies that representing local variability either through $H_t$ or through the volatility $\sigma_t$ mapped via \eqref{eq:IncrVar} is information-equivalent.

\paragraph*{b) An Absolute Benchmark for Market Efficiency}
Because volatility is inherently relative and unrelated to the semimartingale property, it offers no natural basis for assessing market efficiency. By contrast, $H_t$ admits an absolute benchmark: $H_t=1/2$ corresponds exactly to the efficient dynamics of a Brownian semimartingale. Hence the deviation $|H_t - 1/2| \in (0, 1/2)$ provides a direct quantitative measure of market inefficiency and predictability \citep{BPP2015}.

\paragraph*{c) Predictive Content via Mean-Reversion}
Empirically, the time-varying Hurst exponent exhibits stationarity, mean reversion, and an approximately normal distribution centered at $1/2$ -- features consistent with a fractional Ornstein–Uhlenbeck specification for $H_t$ \citep{Cajueiro2004,Peng2018,Garcin2022,Mattera2022}. The observed tendency of financial markets to revert toward the efficient benchmark $H_t=1/2$ motivates the use of the deviation $|H_t - 1/2|$ as an indicator of mean-reversion intensity. This provides a stochastic formalization of the adage “\emph{what goes up must come down}’’ and offers a parsimonious theoretical rationale for reversal phenomena such as those documented by \citep{DeBondt1995}.\\

\begin{table}[ht]
\caption{Financial interpretation of $H_t$}
\label{tab:FinIn}
\footnotesize
\centering
\renewcommand{\arraystretch}{1.3}
\begin{tabularx}{\textwidth}{>{\bfseries}c | X | X | X}
\toprule
$H_t$ & Stochastic properties & Behavioral interpretation & Market implications \\
\midrule
$> \tfrac{1}{2}$ & 
Persistence, smooth paths, $\langle X\rangle_t = 0$ &
New information confirms existing positions &
\textit{Unfair low} volatility, momentum, positive inefficiency, overconfidence, underreaction \\
\midrule
$= \tfrac{1}{2}$ & 
Independence, martingale behavior, $\langle X\rangle_t = t$ (Brownian)&
Information fully incorporated into prices &
\textit{Fair} volatility, sideways market, informational efficiency \\
\midrule
$< \tfrac{1}{2}$ & 
Mean-reversion, rough paths, $\langle X\rangle_t = \infty$ &
New information disrupts existing positions &
\textit{Unfair high} volatility, reversals, negative inefficiency, overreaction \\
\bottomrule
\end{tabularx}
\normalsize
\end{table}

%As summarized in Table \ref{tab:FinIn}, 

The Hurst-H\"{o}lder exponent provides a holistic characterization of market states by discerning both the \textit{magnitude} and \textit{structure} of variability: %It classifies local dynamics into:
%\begin{itemize}\setlength\itemsep{.025cm}
%    \item \textit{Momentum markets} ($H_t > 1/2$): trends and speculative bubbles.
%    \item \textit{Efficient markets} ($H_t \approx 1/2$): sideways movement with minimal predictability.
%    \item \textit{Reversal markets} ($H_t < 1/2$): overreaction and rapid price adjustments.
%\end{itemize}
$H_t$ moves beyond volatility by jointly addressing the ``\textit{how much}'' and ``\textit{how}'' of price variation, offering a unified metric for efficiency, predictability, and regime classification. Moreover, Table \ref{tab:FinIn} summarizes that $H_t$ provides a mathematical basis for integrating rational and behavioral finance within a bounded rationality framework. Rather than being conflicting paradigms, they represent alternating market states, with $H_t$ explicitly quantifying the transition between them. A shift away from the efficient benchmark ($H_t = 1/2$) signals the emergence of systematic irrationality, which the model characterizes as either trend-persistence (bubbles, overconfidence) or anti-persistence (reversals, overreaction), a view that is also consistent with experimental evidence showing that individuals intuitively perceive more irregular dynamics as riskier \citep{Sobolev2016}.

\section{Estimation of the Hurst-H\"{o}lder parameter}\label{sec:Hestimation}
In the previous section, an analytical relationship was established between the Hurst exponent and the standard deviation, both referred to the generic time $t$. Therefore, it is clear that the question becomes how to estimate $H_t$, but detailing the methodologies available for this purpose is beyond the scope of this article. Here we will simply note that the problem of dynamically estimating the time-dependent Hurst--H\"{o}lder exponent has been extensively addressed in the statistical and econometric literature. Numerous methodologies have been proposed for this purpose, including variation-statistical approaches, wavelet-based techniques, and local likelihood methods \citep{IL1997, KW1997, BBCI2000, Coeur2001, Coeur2005, SanchezGraneroetal2012, SanchezGraneroetal2020}. In the empirical implementation, we employ the estimation procedure developed in \citep{PiaBiaPal2018,AngeliniBianchi2023}, to which we refer for the methodological details. This method leverages the local asymptotic self-similarity of a multifractional process $\{X_{i/N}\}_{i=0,\ldots,N}$, which ensures that within a sufficiently small neighborhood $[t-\delta/2,t+\delta/2]$ of any time $t$ a multifractional process of length $N \gg \delta$ behaves like an fBm. Therefore, in a small window of size $\delta$ centered around $t$ the increments are approximately normally distributed with mean zero and variance driven by equation \eqref{eq:varfBm}.
%***************************************************************************
\begin{figure}[ht] 
%\centering
\begin{tikzpicture}[node distance=1cm]

% Row 1
\node[block, text width=3.2cm, anchor=west] (start) at (0,0)
{Data:\\ $X_{i/N},\, i=0,\ldots,N$};
%\node[block, text width=3cm] (start) {Data:\\ $X_{i/n},\, i=0,\ldots,n$};

\node[block, right=1cm of start, text width=5cm] (inc)
  {Compute increments:\\
   $\Delta_i X = X_{t-i/N}-X_{t-(i+1)/N}$};

\node[block, right=1.3cm of inc] (M2)
  {Local variance:\\
   $M_t^{(2)} = \frac1{\delta}\sum |\Delta_iX|^2$};

% Row 2
\node[block, below=1cm of inc, text width=6.2cm] (M2p)
  {Coarse variance:\\
   $M_t^{'(2)} = \frac{2}{\delta}\sum|X_{t-2i/N}-X_{t-2(i+1)/N}|^2$};

\node[block, right=.7cm of M2p] (H2)
  {Unbiased est.:\\
   $\hat H_t^{\delta,N}
   = \frac12\log_2\!\frac{M_t^{'(2)}}{M_t^{(2)}}$};

\node[block, text width=4.5cm, left=.2cm of M2p] (Hb)
  {Biased est.:\\
   $\hat H_t^{\delta,2,C^*,N}
   = \frac{2\log C^*-\log M_t^{(2)}}{2\log(N-1)}$};

% Row 3
\node[block, below=1cm of M2p, text width=6cm,] (bias)
  {Bias estimate:\\
   $h = \frac1{N-\delta+1}\sum
   \left(\hat H_s^{\delta,2,C^*,N}
   -\hat H_t^{\delta,N}\right)$};

\node[block, right=0.8cm of bias] (final)
  {Final corrected:\\
   $\hat H_t^{\delta,2,N}
     = \hat H_t^{\delta,2,C^*,N} - h$};

% Arrows row 1
\draw[arrow] (start) -- (inc);
\draw[arrow] (inc) -- (M2);

% Arrows downward
\draw[arrow] (inc) -- (M2p);
\draw[arrow] (M2) -- (H2);
\draw[arrow] (M2p) -- (H2);
\draw[arrow] (inc) -- (Hb);

% Arrows row 2
\draw[arrow] (Hb) -- (bias);
%\draw[arrow] (M2p) -- (bias);
\draw[arrow] (H2) -- (bias);

% Arrow from bias to final
\draw[arrow] (bias) -- (final);

\end{tikzpicture}
\caption{Pipeline of the estimator of the pointwise Hurst--Hölder exponent $H_t$.} \label{fig:flowchart}
\end{figure}
%***************************************************************************
The approach, summarized in the flowchart of Figure \ref{fig:flowchart}, involves constructing the estimator $\hat{H}_t^{2,\delta,N}$, which merges two estimators: $\hat{H}_t^{\delta,2,C^*,N}$ --- biased (shifted of an unknown scale parameter $C^*$) but with a good rate of convergence --- and $\hat{H}_t^{\delta,N}$ --- unbiased but converging slower. As proved in \citep{B2005b}, this estimator has a rate of convergence $\mathcal{O}(\delta^{-1/2}(\log N)^{-1})$ and
%A key implication of this framework is the establishment of a direct functional relationship between the estimated Hurst exponent and local volatility. In fact, \citep{B2005b,BPP2013} prove that %+

\begin{equation} \label{eq:confint}
    \hat{H}_t^{2,\delta,N}|_{H_t=1/2}\sim \pazccal{N}\left(\frac{1}{2},\frac{1}{2\delta\ln^2(N-1)}\right).
\end{equation}
Table \ref{table:NEFrequencies} reports the nominal and average empirical rejection frequencies, computed over 100 replications of the process, of $\hat{H}_t^{2,\delta,N}\big|{H_t=1/2}$, across different confidence levels, sample sizes (expressed as powers of 2) and values of $\delta$. Table \ref{table:95CI} reproduces the $95\%$-confidence intervals deduced by \eqref{eq:confint}, for different values of $\delta$ and $N$. In the empirical analysis, all estimates are computed using a rolling window of $\delta = 20$ trading days.

\subsection{Goodness-of-fit of the estimator via Monte Carlo simulation}
To evaluate both the accuracy and the robustness of the estimator in recovering time-varying parameters under controlled yet realistic conditions, we have performed a Monte Carlo simulation by employing a highly irregular benchmark function rather than a smooth or purely periodic signal. In the application, we will summarize the Monte Carlo simulation on the individual indices considered in this study.

\begin{table}[htbp] 
\begin{singlespace}
\centering
\caption{Nominal and empirical rejection frequencies (\%) of the estimator across significance levels (average of 100 samples for each entry of the table).}
\scriptsize
\resizebox{\textwidth}{!}{\begin{tabular}{|c|rrrrr|rrrrr|rrrrr|}
\hline
  & \multicolumn{5}{c|}{\textbf{10.0\%}} & \multicolumn{5}{c|}{\textbf{5.0\%}} & \multicolumn{5}{c|}{\textbf{1.0\%}} \\ 
\cline{1-16}
 \diagbox{$\delta$}{$N$} & 1024 & 2048 & 4096 & 8192 & 16384 & 1024 & 2048 & 4096 & 8192 & 16384 & 1024 & 2048 & 4096 & 8192 & 16384 \\
\hline
20  & 10.3 & 10.0 & 10.5 & 10.4 & 10.4 &  5.4 &  5.3 &  5.4 &  5.5 &  5.4 &  1.4 &  1.3 &  1.4 &  1.4 &  1.4 \\
30  & 10.1 &  9.9 & 10.2 & 10.1 & 10.2 &  5.2 &  5.1 &  5.3 &  5.2 &  5.2 &  1.2 &  1.1 &  1.3 &  1.3 &  1.2 \\
40  &  9.9 &  9.7 & 10.1 & 10.1 & 10.2 &  5.1 &  4.9 &  5.2 &  5.2 &  5.1 & 1.2 &  1.0 &  1.3 &  1.2 &  1.1 \\
50  &  9.8 &  9.6 & 10.0 & 10.0 & 10.1 &  5.1 &  4.8 &  5.1 &  5.1 &  5.1 & 1.2 &  1.0 &  1.3 &  1.1 &  1.1 \\
%60  &  9.7 &  9.6 & 10.0 & 10.0 & 10.0 &  5.0 &  4.8 &  5.1 &  5.1 &  5.1 & 1.2 &  1.0 &  1.3 &  1.1 &  1.1 \\
70  &  9.6 &  9.7 & 10.1 & 10.0 & 10.0 &  5.0 &  4.8 &  5.1 &  5.1 &  5.1 & 1.1 &  1.0 &  1.2 &  1.1 &  1.1 \\
%80  &  9.4 &  9.7 & 10.1 & 10.0 & 10.0 &  4.9 &  4.8 &  5.1 &  5.1 &  5.0 & 1.1 &  1.0 &  1.2 &  1.1 &  1.1 \\
%90  &  9.5 &  9.6 & 10.1 & 10.0 & 10.0 &  4.9 &  4.7 &  5.2 &  5.0 &  5.0 & 1.1 &  0.9 &  1.2 &  1.1 &  1.0 \\
100 &  9.6 &  9.6 & 10.2 &  9.9 &  9.9 &  4.9 &  4.7 &  5.2 &  5.0 &  5.0 & 1.1 &  0.9 &  1.2 &  1.1 &  1.0 \\
%120 &  9.8 &  9.5 & 10.3 &  9.8 &  9.9 &  4.9 &  4.7 &  5.3 &  4.9 &  5.0 & 1.2 &  1.0 &  1.3 &  1.0 &  1.0 \\
%140 & 10.0 &  9.4 & 10.3 &  9.8 &  9.9 &  5.1 &  4.7 &  5.4 &  4.9 &  4.9 & 1.1 &  0.9 &  1.3 &  1.0 &  0.9 \\
150 & 10.0 &  9.5 & 10.3 &  9.8 &  9.9 &  4.9 &  4.6 &  5.5 &  4.9 &  4.9 & 1.1 &  0.9 &  1.3 &  1.0 &  0.9 \\
%180 &  9.9 &  9.5 & 10.4 &  9.9 &  9.9 &  4.9 &  4.6 &  5.5 &  4.9 &  4.9 & 1.1 &  0.9 &  1.3 &  1.0 &  0.9 \\
200 &  9.9 &  9.5 & 10.4 &  9.9 &  9.8 &  4.9 &  4.5 &  5.5 &  5.0 &  4.8 & 1.1 &  0.9 &  1.3 &  1.0 &  0.9 \\
%250 &  9.8 &  9.1 & 10.4 &  9.8 &  9.6 &  5.2 &  4.5 &  5.6 &  5.0 &  4.8 & 1.3 &  0.8 &  1.3 &  1.0 &  0.9 \\
300 & 10.2 &  9.0 & 10.2 &  9.8 &  9.5 &  5.5 &  4.3 &  5.3 &  4.9 &  4.7 & 1.4 &  0.8 &  1.3 &  1.1 &  0.9 \\
\hline
\end{tabular}
}
\label{table:NEFrequencies}
\end{singlespace}
\end{table}

\begin{table}[htbp]
\begin{singlespace}
\centering
\caption{95\% Confidence Intervals $\big(\mu \pm 1.96\sqrt{1/(2\delta\ln^2(N-1))}\big)$ for $\mu = 1/2$.}
\scriptsize{
\setlength{\tabcolsep}{4pt}
\renewcommand{\arraystretch}{1.15}
\begin{tabular}{|c|ccccccc|}
\hline
\diagbox{$\delta$}{$N$} & 1000 & 2000 & 3000 & 4000 & 5000 & 7000 & 10000 \\
\hline
10  & (0.444, 0.556) & (0.454, 0.546) & (0.458, 0.542) & (0.460, 0.540) & (0.462, 0.538) & (0.464, 0.536) &  (0.467, 0.533) \\
20  & (0.472, 0.528) & (0.477, 0.523) & (0.479, 0.521) & (0.480, 0.520) & (0.481, 0.519) & (0.483, 0.517) & (0.484, 0.516) \\
30  & (0.480, 0.520) & (0.483, 0.517) & (0.484, 0.516) & (0.485, 0.515) & (0.486, 0.514) & (0.487, 0.513) &  (0.488, 0.512) \\
40  & (0.484, 0.516) & (0.486, 0.514) & (0.487, 0.513) & (0.487, 0.513) & (0.488, 0.512) & (0.489, 0.511) & (0.489, 0.511) \\
50  & (0.487, 0.513) & (0.488, 0.512) & (0.488, 0.512) & (0.489, 0.511) & (0.489, 0.511) & (0.490, 0.510) & (0.490, 0.510) \\
60  & (0.489, 0.511) & (0.489, 0.511) & (0.489, 0.511) & (0.490, 0.510) & (0.490, 0.510) & (0.490, 0.510) & (0.490, 0.510) \\
70  & (0.490, 0.510) & (0.490, 0.510) & (0.490, 0.510) & (0.490, 0.510) & (0.490, 0.510) & (0.490, 0.510) & (0.491, 0.509) \\
80  & (0.490, 0.510) & (0.490, 0.510) & (0.490, 0.510) & (0.491, 0.509) & (0.491, 0.509) & (0.491, 0.509) & (0.491, 0.509) \\
90  & (0.491, 0.509) & (0.491, 0.509) & (0.491, 0.509) & (0.491, 0.509) & (0.491, 0.509) & (0.491, 0.509) & (0.491, 0.509) \\
100 & (0.491, 0.509) & (0.491, 0.509) & (0.491, 0.509) & (0.491, 0.509) & (0.491, 0.509) & (0.491, 0.509) & (0.491, 0.509) \\
\hline
\end{tabular}
}
\label{table:95CI}
\end{singlespace}
\end{table}

\begin{figure}[t]%[htbp]
    \begin{minipage}{0.49\textwidth}
        \centering
        \captionsetup{margin=.9cm}
        \includegraphics[width=\linewidth]{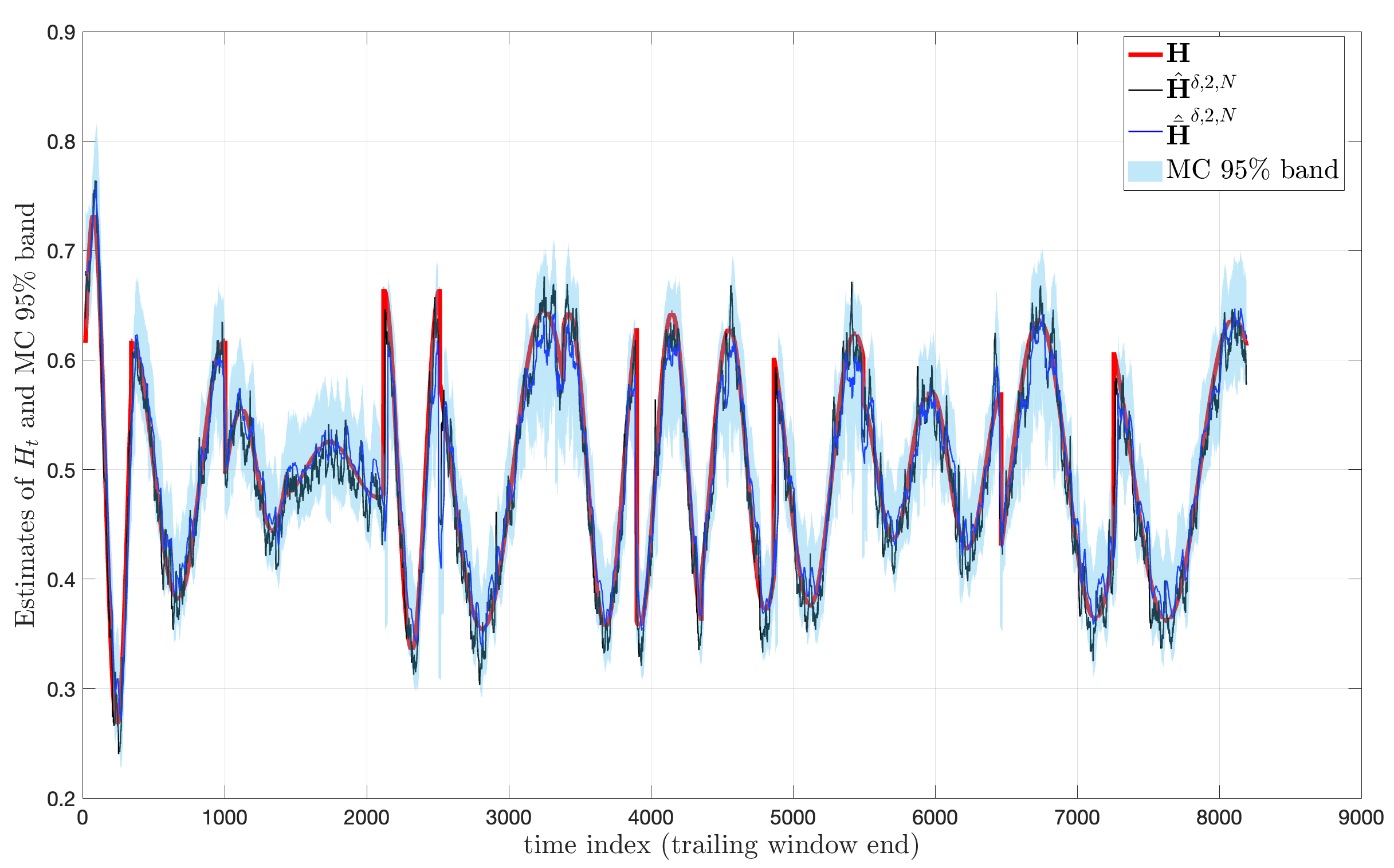}
        \caption{Original vs Monte Carlo re-estimation with $95\%$ MC band.} \label{fig:Sim_H_est_MC} 
    \end{minipage} 
    \begin{minipage}{0.53\textwidth}
        \centering
        \captionsetup{margin=.8cm}
        %[width=\linewidth]
        \includegraphics[height=0.22\textheight,keepaspectratio]{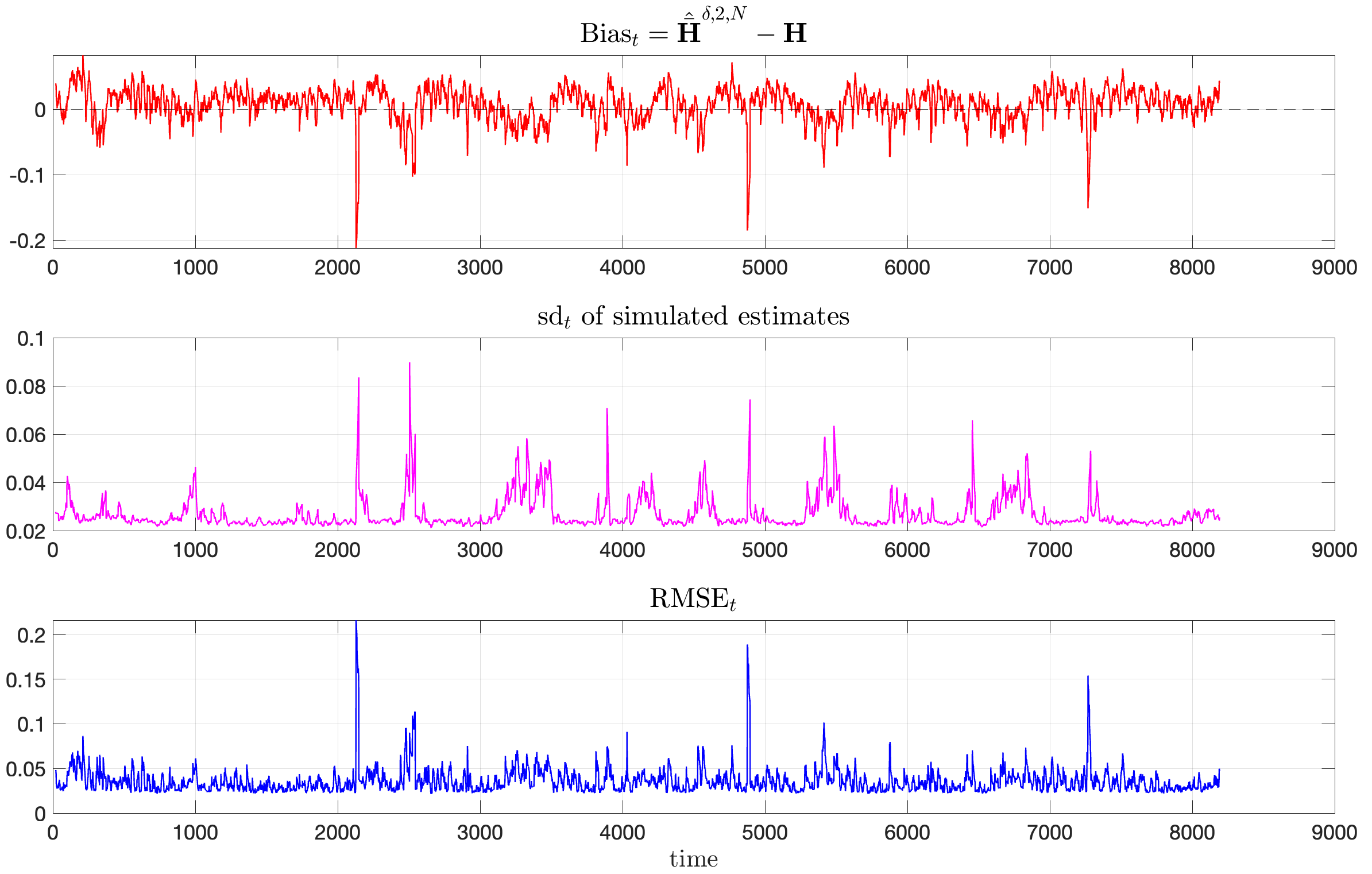}
        \caption{Bias, Standard Deviation and Root Mean Square Error.} \label{fig:Sim_Bias}
    \end{minipage}
\end{figure}

Specifically, we generate a piecewise-sinusoidal process whose amplitude, phase, and local frequency vary randomly across cycles, producing an oscillatory pattern that departs from classical harmonic regularity (e.g., red curve in Figure \ref{fig:Sim_H_est_MC}). Such a construction serves as a stringent stress test: it preserves the overall smoothness of a sinusoidal waveform while introducing local nonstationarities, abrupt changes in oscillation length, and uneven amplitude modulations. An estimator capable of accurately recovering this type of signal can be expected to perform reliably even in empirical contexts where the underlying dynamics are far from idealized or stationary. %This “irregular sinusoid” provides a controlled yet challenging setting for assessing both the flexibility and the stability of the estimation procedure.

Let $N$ be the total length and $C$ be the number of cycles. For each cycle $c = 1,\dots,C$, let $A_c > 0$ denote its amplitude, $\phi_c \in [0,2\pi)$ its phase, and $L_c \in \mathbb{N}$ its length in samples, such that $\sum_{c=1}^{C} L_c = N$. Define the cumulative boundaries 
\begin{equation*}
    n_0 = 0, \qquad 
n_c = \sum_{j=1}^{c} L_j, \qquad c=1,\dots,C,
\end{equation*}
and set the mean level $\mu = 0.5$ (benchmark for an efficient market). Then, for discrete time indices $t = 1,\dots,N$, the signal is defined as
\begin{equation*}
H_{t} \;=\; \mu \;+\; A_{c(t)} \,\sin\!\left(
2\pi\,\frac{t - n_{c(t)-1}}{L_{c(t)}} + \phi_{c(t)}
\right),
\end{equation*}
where $c(t)$ is the unique cycle index satisfying 
$n_{c(t)-1} < t \le n_{c(t)}$. To introduce variability in amplitude, phase, and period, we define
\begin{equation*}
A_c = A_0 + \sigma_A\,\varepsilon_c, 
\qquad 
\varepsilon_c \sim \pazccal{N}(0,1),
\end{equation*}
\begin{equation*}
\phi_c \sim \mathrm{Unif}[0,2\pi),
\end{equation*}
\begin{equation*}
L_c \propto \frac{N}{C}\,\bigl(1 + \rho\,\eta_c\bigr),
\qquad 
\eta_c \sim \pazccal{N}(0,1),
\end{equation*}
where $A_0 > 0$ is the base amplitude, $\sigma_A \ge 0$ controls its variability, 
and $\rho \in (0,1)$ controls the variability of cycle lengths. 
The lengths are finally rescaled so that $\sum_c L_c = N$, and bounded below by $L_{\min}$ to avoid degenerate cycles.\\ %An example of "wild" $H_{k}$ with $N=8,192$ is provided in Figure \ref{fig:Wild_f}.
%\begin{figure}[ht]
%\centering
%\captionsetup{margin=1.33cm}
%\includegraphics[scale=.15, trim=0pt 0pt 0pt 0pt]{images/Sim_Hk_rit.png}
%\caption{Randomized sinusoidal-like signal around 0.5 (11 cycles)}
%\label{fig:Wild_f} 
%\end{figure}
The Monte Carlo simulation consists of the following five steps.
\begin{enumerate}[leftmargin=*] \setlength\itemsep{.9pt}
    \item Function $\mathbf{H}=(H_t)_{t=1,\ldots,N}$ is taken as input to simulate a path of MPRE\footnote{The MPRE was simulated using the MATLAB function \textit{mBmQuantifKrigeage()} from the FracLab Toolbox 2.2, released by INRIA at https://project.inria.fr/fraclab/, which is based on the Wood and Chan circulant matrix
method \citep{ChanWood98}.}, $\mathbf{X}=(X_t)_{t=1,\ldots,N}$ (red curve in Figure \ref{fig:Sim_H_est_MC}).
    \item $\mathbf{X}$ is used to estimate $\hat{\mathbf{H}}^{\delta,2,N}=(\hat{H}^{\delta,2,N}_{t})_{t=\delta,\ldots,N}$ (black curve in Figure \ref{fig:Sim_H_est_MC}).
    \item The estimated $\hat{\mathbf{H}}^{\delta,2,N}$ is treated as the ``true'' functional parameter path and used to generate $R$ independent synthetic trajectories $\mathbf{X}^{(r)} = (X^{(r)}_{t})_{t=\delta,\ldots,N}$, for $r=1,\dots,R$, through the data-generating process assumed by the model.
    \item The estimator is re-applied to each simulated trajectory, 
yielding $R$ corresponding sequences $\hat{\mathbf{H}}^{\delta,2,N}(r)=(\hat{H}^{\delta,2,N}_{t}(r))_{t=2\delta,\ldots,N}$. The blue curve in Figure \ref{fig:Sim_H_est_MC} represents the average $\hat{\bar{\mathbf{H}}}^{\delta,2,N}=\left(\frac{1}{R}\sum_{r=1}^R \hat{H}^{\delta,2,N}_t(r)\right)_{t=2\delta,\ldots,N}$ across all simulations.
    \item Finally, the set of replications $\{\hat{H}^{\delta,2,N}_k{(r)}\}_{r=1,\ldots,R;\;k=\delta,\ldots,N}$ is compared with the reference path to quantify bias, dispersion, and coverage at each step $k$ as well as global metrics of accuracy.
\end{enumerate}
\begin{figure}%[hbp]
\centering
% -------- LEFT: FIGURE (0.60\textwidth) --------
\begin{minipage}[t]{0.60\textwidth}
  \vspace{0pt}%
  \centering
  % Constrain caption to the minipage width (prevents overflow)
  %\captionsetup{width=\linewidth}
  \captionsetup{margin=.8cm}
  \includegraphics[width=\linewidth,height=0.25\textheight,keepaspectratio]{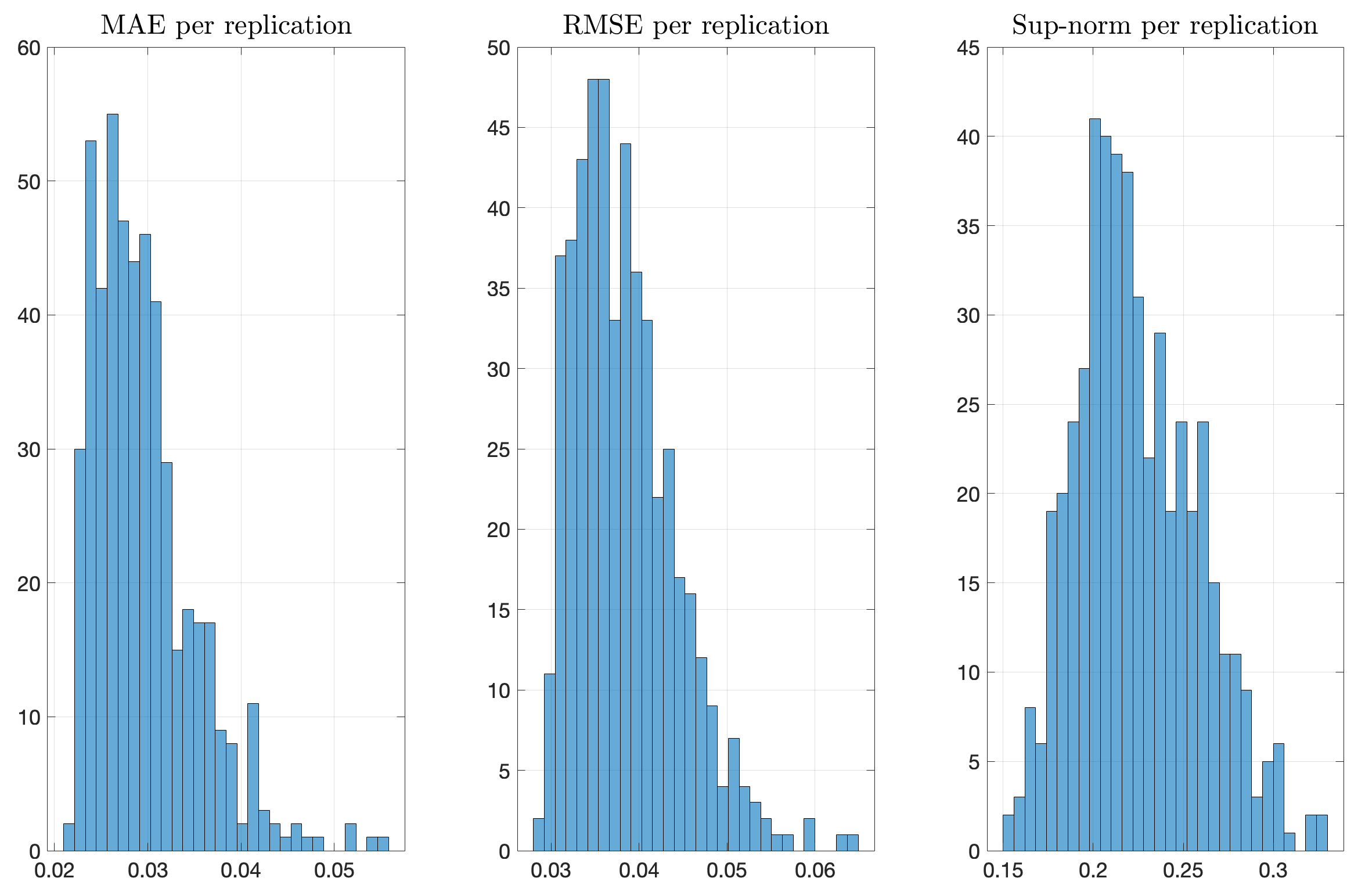}
  \caption{Distributions of MAE, RMSE and Sup-norm (500 replications).}
  \label{fig:Sim_MAE_RMSE_SUP}
\end{minipage}%
\hfill
% -------- RIGHT: TABLE (0.37\textwidth) --------
\begin{minipage}[t]{0.37\textwidth}
  \vspace{0pt}%
  \centering
  \captionsetup{width=\linewidth}
  \scriptsize
  \captionof{table}{Monte Carlo summary statistics of estimator performance.}
  \label{tab:MC_Sim_summary_values}
  \begin{tabular}{@{}lc@{}}
    \toprule
    \textbf{Statistic} & \textbf{Value} \\
    \midrule
    MAE\_mean        & 0.02958 \\
    MAE\_median      & 0.02858 \\
    MAE\_p95         & 0.04059 \\
    RMSE\_mean       & 0.03837 \\
    RMSE\_median     & 0.03743 \\
    RMSE\_p95        & 0.04921 \\
    SUP\_mean        & 0.22405  \\
    SUP\_median      & 0.21883  \\
    SUP\_p95         & 0.28406  \\
    CoverageMeanPct  & 95.2\%   \\
    \bottomrule
  \end{tabular}
\end{minipage}
\end{figure}
 
\paragraph{Evaluation criteria}
For each fixed time, the Monte Carlo distribution of the re-estimated values provides an empirical measure of uncertainty around the original estimate. From these distributions, we compute, pointwise in $t$, the Monte Carlo bias $\mathrm{Bias}_t = \hat{\bar{\mathbf{H}}}^{\delta,2,N} - \mathbf{H}$, the standard deviation $\mathrm{sd}_t$, and the root-mean-square error $\mathrm{RMSE}_t = \sqrt{\mathrm{Bias}_t^2 + \mathrm{sd}_t^2}$.
Empirical $95\%$ confidence bands are obtained from the quantiles of 
$\{\hat{H}^{\delta,2,N}_t(r)\}$, and their coverage rate—i.e., the proportion of replications for which the original $\hat{H}_t^{\delta,2,N}$ lies within the band—is used to assess calibration quality.
At a global level, aggregated measures such as the mean absolute error (MAE), the average RMSE, and the maximum absolute deviation (sup-norm) across $k$ summarize the overall agreement between simulated and original estimates. Figures \ref{fig:Sim_H_est_MC}, \ref{fig:Sim_Bias} and \ref{fig:Sim_MAE_RMSE_SUP}, and Table \ref{tab:MC_Sim_summary_values} summarize the Monte Carlo study (the table reports mean, median, and 95\textsuperscript{th} percentile 
for the absolute error (MAE), root-mean-square error (RMSE), and maximum deviation (SUP), together with the average coverage rate of the 95\% confidence bands).
%\medskip
%\noindent
%\textbf{Continuous-time analogue.}
%For a continuous domain $t \in [0,1]$, let $0 = t_0 < t_1 < \dots < t_C = 1$ be a partition with durations $T_c = t_c - t_{c-1}$.
%Then
%\[
%f(t) 
%\;=\; 
%\mu 
%\;+\;
%A_{c(t)} 
%\,\sin\!\left(
%2\pi\,\frac{t - t_{c(t)-1}}{T_{c(t)}} + \phi_{c(t)}
%\right),
%\qquad 
%t \in (0,1],
%\]
%where $c(t)$ is the unique index such that $t_{c(t)-1} < t \le t_{c(t)}$.
%Sampling at $t_k = k/N$ yields the discrete version above with $L_c \approx N T_c$.
%The resulting process is a piecewise-sinusoidal signal centered at $\mu = 0.5$, with cycle-specific amplitudes, phases and lengths. The randomness in $(A_c, L_c, \phi_c)$ introduces heterogeneous oscillations while preserving the sinusoidal shape within each cycle. Continuity at boundaries can be enforced, if desired, by choosing the phases  $\phi_c$ so that $f(n_{c-1}) = f(n_{c-1}+1)$ (or by matching slopes), but in this work $\phi_c \sim \mathrm{Unif}[0,2\pi)$ is retained to allow mild irregularities consistent with the simulation design.
%********************************************************

\section{Empirical analysis} \label{sec:Application}
In this section, we will apply the methodology presented in the preceding sections to the set of 14 daily global stock indices summarized in Table \ref{tab:Dataset}. 

We highlight that while the empirical analysis in this study relies on daily observations, extending the proposed framework to high-frequency (intraday) data presents both distinct opportunities and methodological challenges. Theoretically, higher frequencies yield larger sample sizes over shorter calendar windows, potentially improving the precision of local regularity estimators. However, at very fine time scales, the observed price process becomes heavily contaminated by market microstructure noise, such as bid-ask bounce \citep{roll1984simple}, asynchronous trading \citep{epps1979comovements}, and discrete price increments \citep{ait2005often}. This noise systematically biases the estimated Hurst–Hölder exponent downwards, imposing a spurious artificial roughness (anti-persistence) that reflects frictional trading mechanics rather than the true informational geometry of the market. Consequently, directly applying the fair volatility benchmark to tick-by-tick or 1-minute data would likely overestimate the degree of market inefficiency. Therefore, adapting this methodology to an intraday setting requires a careful, asset-specific calibration of the sampling frequency to achieve an optimal trade-off between maximizing data points and filtering out microstructure-induced distortions. Given these structural constraints, alongside the practical unavailability of decades-long, high-quality intraday datasets required to analyze long-term volatility cycles without small-sample bias, the daily frequency serves as the most robust baseline for evaluating the structural properties of fair volatility.

\begin{table}[htbp]
\centering
\scriptsize
\caption{Dataset}
\label{tab:Dataset}
\begin{tabular}{@{}lllccr@{}}
\toprule
\textbf{Index} & \textbf{Country} & \textbf{Ticker} & \textbf{Start Date} & \textbf{End Date} & \textbf{Size $(N)$} \\
\midrule
S\&P 500 Index                        & USA (USA)      & SPX    & 1927-12-30 & 2025-08-22 & 24,527 \\
NASDAQ Composite Index                & USA (USA)      & CCMP   & 1971-02-05 & 2025-08-22 & 13,753 \\
EURO STOXX 50 & EUR (EUR)       & SX5E   & 2007-03-30 & 2025-08-22 & 4,612  \\
FTSE 100 Index                        & United Kingdom (GBR)        & UKX    & 1984-01-03 & 2025-08-22 & 10,519 \\
Hang Seng Index                       & Hong Kong (HKG) & HSI    & 1986-12-31 & 2025-08-22 & 9,538  \\
FTSE Bursa Malaysia KLCI Index        & Malaysia (MYS)  & KLCI   & 1993-12-03 & 2025-08-22 & 7,795  \\
KOSPI Index                           & South Korea (KOR)     & KOSPI  & 1996-12-11 & 2025-08-22 & 7,067  \\
MOEX Russia Index             & Russia (RUS)   & IMOEX  & 2013-03-05 & 2024-06-14 & 2,795  \\
NIFTY 50 Index                        & India (IND)    & NIFTY  & 2007-09-17 & 2025-08-22 & 4,399  \\
Nikkei 225                            & Japan (JPN)    & NKI    & 1965-01-05 & 2025-08-22 & 14,909 \\
SET Index (Thailand)                  & Thailand (THA) & SET    & 1996-12-11 & 2025-08-22 & 7,000  \\
Shanghai Composite Index              & Shanghai (CHN) & SHCOMP & 1997-07-02 & 2025-08-22 & 6,819  \\
FTSE Straits Times Index              & Singapore (SGP) & FSSTI  & 1987-12-28 & 2025-08-22 & 9,408  \\
Taiwan Stock Exchange Weighted Index  & Taiwan (TWN)  & TWSE   & 1997-07-02 & 2025-08-22 & 6,900  \\
\bottomrule
\end{tabular}
\end{table}

We assume that the pricing process can be represented as an MPRE—yet general framework capable of encompassing a wide range of models used to describe financial dynamics. Under this assumption, the empirical procedure proceeds as follows:
\begin{enumerate}
    \item Estimate $\hat{\mathbf{H}}^{2,\delta,N}$ from the data using the estimator described in Section \ref{sec:Hestimation}, with a rolling-window length of $\delta=20$ trading days.%Estimation of local regularity.
    \item Perform the Monte Carlo simulation described in the previous Section from step 2 to 5 (with 500 replications), where $\mathbf{X}$ is the log-index sequence. Table \ref{table:metrics} summarizes MAE, RMSE and SUP along with their 95th percentiles.
    \item Compute the historical volatility $\hat\sigma_t^{\text{hist}}$ of returns by applying the standard deviation formula over the same rolling window $\delta$ used in the previous step.%Estimation of empirical historical volatility. 
    \item Using the previous estimations $\hat{\mathbf{H}}^{2,\delta,N}$and $\hat\sigma_t^{\text{hist}}$, estimate the scale parameter $\hat\nu_t$ by inverting the relation \eqref{eq:IncrVar}
    with $h=1/N$.
    \item Simulate $100$ trajectories of MPRE, with the estimated $\hat{\mathbf{H}}^{2,\delta,N}$ as the regularity function, a unit time support and a final unit standard deviation as prescribed by the relation \eqref{eq:IncrVar}, that is $\hat\nu_0\sqrt{A_{\hat H_0}}$. %Setting the scale function $\nu_1=\text{sd}(\Delta_1\mathbf{X})$,
    Compute the standard deviation of MPRE increments — hereafter referred to as the theoretical volatility — following the procedure as in Step 3.%Simulation of MPRE paths. 
    %\item Using the asymptotic relation \eqref{eq:IncrVar} with increments $h=1/N$, estimate the scale function $\nu_t$ from the theoretical volatility. The resulting values slightly deviate from the constant 1 due to numerical approximation.
    %\item Rescaling. Normalize the theoretical volatility by dividing it by the estimated $\nu_t$, thereby imposing a unit scale function.%Estimation of the scale function. 
    \item Compare the historical volatility with this rescaled theoretical volatility to validate the model.%Model validation and comparison. 
\end{enumerate}

This comparison serves two primary purposes:
\begin{itemize}
    \item[(i)] \textit{Model validation}: a close correspondence between the two volatility measures provides empirical support for the MPRE as a suitable model for asset-price dynamics.
    \item[(ii)] \textit{Interpretation within a benchmark framework}: volatility can be expressed relative to the benchmark that corresponds to the value expected under market efficiency, i.e., when the price process follows a semimartingale. Concretely, when transformed through relation \eqref{eq:IncrVar}, the confidence interval associated with $H_t=1/2$ (see equation \eqref{eq:confint}) yields the bounds of the confidence interval for volatility. This allows us to interpret volatility values within the interval as \textit{fair}, while values outside it may be regarded as \textit{unfairly low} or \textit{unfairly high}, in line with the description provided in Table \ref{tab:FinIn}.
\end{itemize}

%Before reporting the empirical findings, we clarify the rationale for the assumption 
%$\nu_t=1$ used in the simulation step. For simplicity, the scale function is fixed at $\nu_t=1$ and this parsimonious choice allows us to test whether the time-varying regularity $H_t$ alone is sufficient to capture the empirical variability of financial prices. Deviations between theoretical and historical volatility can then be attributed to local fluctuations in the scale function. These could enhance the ability of the model to account for jumps or discontinuities that in this version are still not considered.\\

\begin{table}[ht]
\centering
\caption{Summary metrics (MAE, RMSE, SUP): mean, median, 95th percentile (p95).}
\label{table:metrics}
\scriptsize
\begin{tabular}{lrrrrrrrrr}
\toprule
& \multicolumn{3}{c}{MAE} & \multicolumn{3}{c}{RMSE} & \multicolumn{3}{c}{SUP} \\
\cmidrule(lr){2-4}\cmidrule(lr){5-7}\cmidrule(lr){8-10}
& mean & median & p95 & mean & median & p95 & mean & median & p95 \\
\midrule
SPX    & 0.03069 & 0.02979 & 0.03965 & 0.03934 & 0.03860 & 0.04919 & 0.19065 & 0.18908 & 0.23406 \\
CCMP   & 0.03246 & 0.03207 & 0.04081 & 0.04240 & 0.04220 & 0.05263 & 0.20712 & 0.20804 & 0.25833 \\
SX5E   & 0.03937 & 0.03661 & 0.06182 & 0.04704 & 0.04446 & 0.06896 & 0.15688 & 0.15447 & 0.19395 \\
UKX    & 0.02812 & 0.02695 & 0.03705 & 0.03596 & 0.03472 & 0.04565 & 0.17143 & 0.17091 & 0.21613 \\
HSI    & 0.03449 & 0.03383 & 0.04620 & 0.04454 & 0.04414 & 0.05699 & 0.20909 & 0.20622 & 0.25879 \\
KLCI   & 0.04576 & 0.04515 & 0.06317 & 0.05716 & 0.05659 & 0.07556 & 0.20173 & 0.20061 & 0.24352 \\
KOSPI  & 0.02948 & 0.02812 & 0.03877 & 0.03730 & 0.03623 & 0.04723 & 0.15862 & 0.15649 & 0.20065 \\
IMOEX  & 0.03575 & 0.03268 & 0.05424 & 0.04500 & 0.04241 & 0.06349 & 0.18475 & 0.18362 & 0.22520 \\
NIFTY  & 0.03257 & 0.03004 & 0.04858 & 0.04133 & 0.03899 & 0.05708 & 0.21079 & 0.20865 & 0.26586 \\
NKI    & 0.02952 & 0.02866 & 0.03794 & 0.03891 & 0.03799 & 0.04898 & 0.20908 & 0.20181 & 0.28068 \\
SET    & 0.03590 & 0.03451 & 0.04950 & 0.04608 & 0.04529 & 0.06147 & 0.18723 & 0.18606 & 0.23218 \\
SHCOMP & 0.04013 & 0.03870 & 0.05643 & 0.05111 & 0.05046 & 0.06809 & 0.23776 & 0.23675 & 0.30993 \\
FSSTI  & 0.03130 & 0.03049 & 0.04072 & 0.04092 & 0.03993 & 0.05113 & 0.20905 & 0.20665 & 0.26039 \\
TWSE   & 0.02840 & 0.02750 & 0.03687 & 0.03621 & 0.03516 & 0.04543 & 0.14829 & 0.14608 & 0.18412 \\
\bottomrule
\end{tabular}
\end{table}

Figures \ref{fig:SPX}–\ref{fig:TWSE} present, for each index, three panels.
The top panel shows the estimated parameter $H_t$ together with its confidence interval around the value $1/2$, which represents market efficiency; the green shaded area marks this interval.
The middle panel compares the theoretical volatility (in red), computed following steps 3 and 5, with the empirical historical volatility (in blue), estimated as described in step 2. The green region again indicates the 95\% confidence interval for fair volatility. This interval is constructed using in equation \eqref{eq:IncrVar} the values $H=\frac{1}{2}\pm z_{1-\alpha/2}\sqrt{\frac{1}{2\delta\ln^2(N-1)}}$.
The bottom panel displays the function $\nu_t$, estimated according to step 4.
\newpage
\begin{landscape}
\begin{singlespace}
\setcounter{page}{17}
\begin{minipage}[t]{0.875\textwidth} % Left area (table)
\begin{table}[H]
\centering
\caption{Summary statistics of estimates.}
\label{tab:hurst_vol_all}
\resizebox{\linewidth}{!}{
%\resizebox{\textwidth}{!}{%
\begin{tabular}{lccccccc} 
\toprule
\textbf{Index} & \textbf{SPX} & \textbf{CCMP} & \textbf{SX5E} & \textbf{UKX} & \textbf{HSI} & \textbf{KLCI} & \textbf{KOSPI} \\
\midrule
\multicolumn{8}{l}{\textbf{Hurst-H\"{o}lder parameter}} \\
Mean   & 0.4735 & 0.5105 & 0.5294 & 0.5150 & 0.5039 & 0.5579 & 0.5168 \\
St.Dev & 0.0520 & 0.0563 & 0.0542 & 0.0460 & 0.0486 & 0.0651 & 0.0608 \\
Range  & 0.3507 & 0.3528 & 0.3297 & 0.3197 & 0.3490 & 0.4451 & 0.3372 \\
Kurtosis & 3.7679 & 3.1850 & 3.4356 & 4.1589 & 4.4772 & 4.0575 & 2.6381 \\
Skewness & -0.6909 & -0.5508 & -0.5126 & -0.8275 & -0.8965 & -0.7992 & -0.4438 \\
$95\%$ Confidence interval & [0.469,0.531] & [0.467,0.533] & [0.463,0.537] & [0.467,0.533] & [0.466,0.534] & [0.465,0.535] & [0.465,0.535]  \\
\multicolumn{8}{l}{\textbf{ADF test}} \\
\hspace{5em} pValue &  0.001 & 0.001 & 0.001 & 0.001 & 0.001 & 0.001 & 0.001  \\
\hspace{5em} Stat &  -10.459 & -7.908  &  -5.823 & -7.783  &  -7.712 & -6.201 & -6.166 \\
\hspace{5em} cValue & -3.412 & -3.412 &  -3.414 & -3.412 & -3.412 &  -3.413 & -3.413  \\
\midrule
\multicolumn{8}{l}{\textbf{Historical volatility}} \\
Mean   & 0.0098 & 0.0105 & 0.0123 & 0.0095 & 0.0139 & 0.0089 & 0.0139 \\
St.Dev  & 0.0069 & 0.0071 & 0.0067 & 0.0052 & 0.0086 & 0.0079 & 0.0086 \\
Range   & 0.0647 & 0.0580 & 0.0486 & 0.0481 & 0.0944 & 0.0969 & 0.0543\\
Kurtosis  & 16.0775 & 11.5115 & 10.1863 & 17.8791 & 45.1439 & 74.2495 & 6.3050 \\
Skewness    & 3.0029 & 2.4622 & 2.2375 & 3.0902 & 3.9255 & 5.1524 & 1.7096  \\
$95\%$ C.I. fair volatility & [0.005,0.009] & [0.006,0.012] & [0.010,0.021] & [0.007,0.014] & [0.007,0.014] & [0.008,0.016] & [0.009,0.017] \\
\midrule
       &  &  &  &  &  &  &   \\
       &  &  &  &  &  &  &   \\
\midrule
\textbf{Index} & \textbf{IMOEX} & \textbf{NIFTY} & \textbf{NKI} & \textbf{SET} & \textbf{SHCOMP} & \textbf{FSSTI} & \textbf{TWSE} \\
\midrule
\multicolumn{8}{l}{\textbf{Hurst-H\"{o}lder parameter}} \\
Mean   & 0.5524 & 0.5448 & 0.4741 & 0.5184 & 0.5153 & 0.5216 & 0.502 \\
St.Dev & 0.0589 & 0.0565 & 0.0516 & 0.0613 & 0.0539 & 0.0528 & 0.0525 \\
Range & 0.4427 & 0.3452 & 0.3769 & 0.3712 & 0.3136 & 0.3282 & 0.2942 \\
Kurtosis & 7.3254 & 4.0534 & 3.1413 & 2.9688 & 2.7653 & 3.3130 & 2.4866 \\
Skewness  & -1.3781 & -0.9117 & -0.1307 & -0.2557 & -0.2764 & -0.6882 & -0.2983 \\
$95\%$ Confidence interval & [0.461,0.539] & [0.463,0.537] & [0.468,0.532] & [0.465,0.535] & [0.465,0.535] & [0.466,0.534] & [0.465,0.535] \\
\multicolumn{8}{l}{\textbf{ADF test}} \\
\hspace{5em} pValue & 0.004 & 0.001 & 0.001 & 0.001 & 0.001 & 0.001 & 0.001 \\
\hspace{5em} Stat & -4.274& -5.227  & -9.735 & -6.562 & -6.748 & -7.591   & -6.790  \\
\hspace{5em} cValue &  -3.414 & -3.414 & -3.412 & -3.413 & -3.413 & -3.412 &  -3.413 \\
\midrule
\multicolumn{8}{l}{\textbf{Historical volatility}} \\
Mean  & 0.0115 & 0.0111 & 0.0112 & 0.0121 & 0.0131 & 0.0098 & 0.0121 \\
St.Dev & 0.0099 & 0.0071 & 0.0063 & 0.0074 & 0.0068 & 0.0058 & 0.0061 \\
Range  & 0.1030 & 0.0532 & 0.0711 & 0.0510 & 0.0395 & 0.0525 & 0.0360 \\
Kurtosis & 59.3325 & 14.0777 & 14.5867 & 7.8507 & 4.8391 & 11.4568 & 4.8389 \\
Skewness  & 6.6030 & 2.8786 & 2.3819 & 1.9074 & 1.4123 & 2.3949 & 1.3233 \\
$95\%$ C.I. fair volatility & [0.013,0.027] & [0.011,0.021] & [0.006,0.012] & [0.009,0.017] & [0.009,0.017] & [0.007,0.015] & [0.009,0.017] \\
\bottomrule
\end{tabular}
}
\end{table}
\end{minipage}
\hfill
\begin{minipage}[t]{0.6\textwidth}
    \begin{figure}[H]
        \centering
\includegraphics[width=\linewidth,height=0.45\textheight,keepaspectratio]{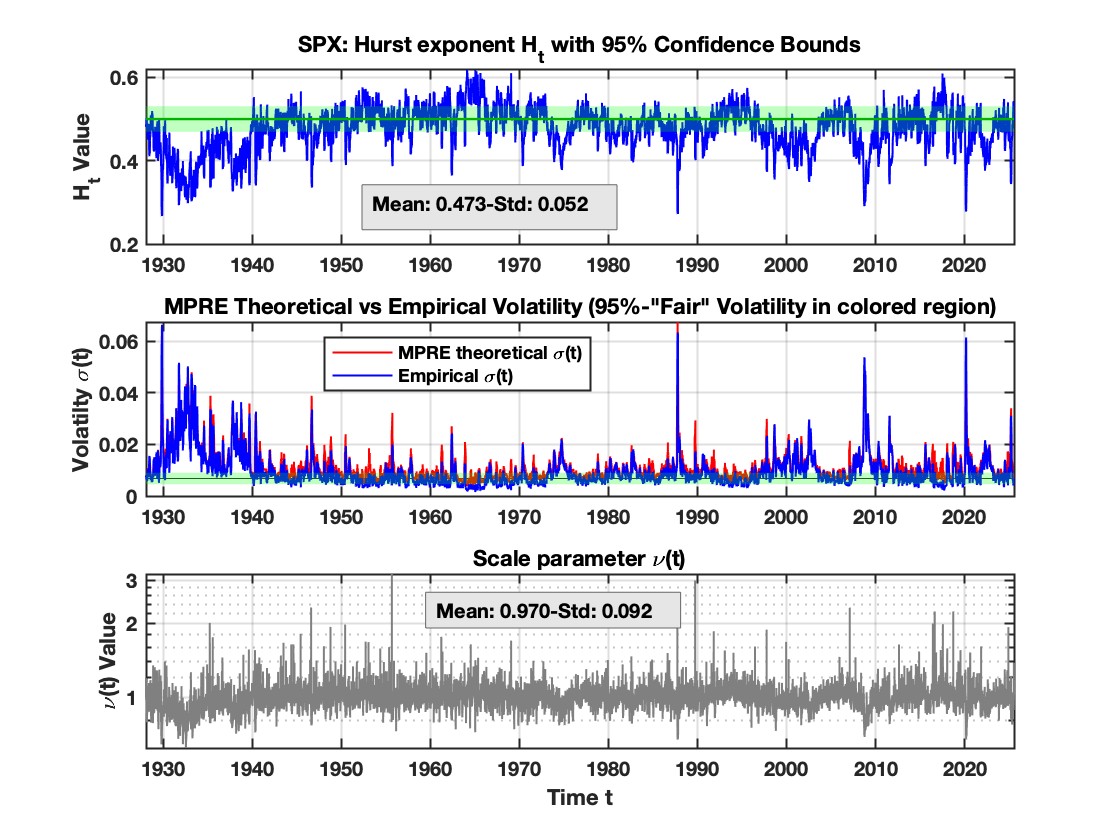}
       \caption{SPX Index} \label{fig:SPX}
    \end{figure}
    %\vspace{0.02\textheight} % small gap
\vspace{-1cm}
    \begin{figure}[H]
        \centering
\includegraphics[width=\textwidth,height=0.45\textheight,keepaspectratio]{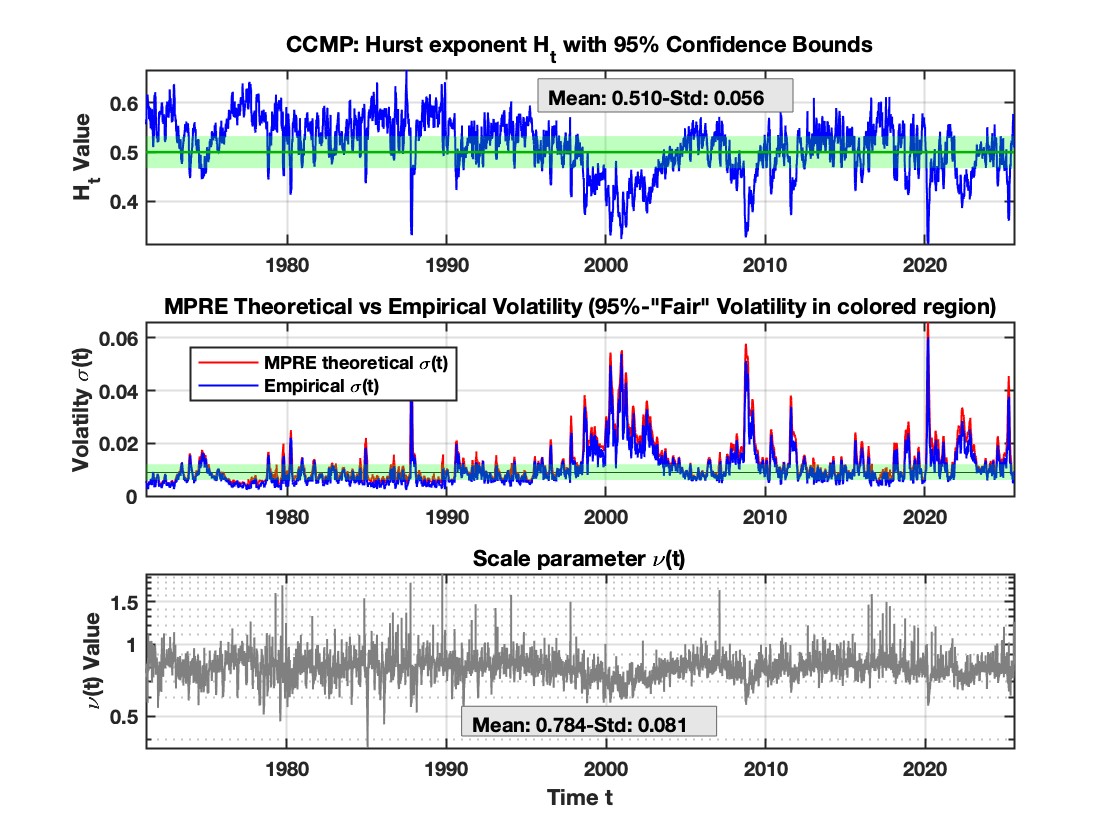}
        \caption{CCMP Index} \label{fig:CCMP}
    \end{figure}
\end{minipage}
\end{singlespace}
\end{landscape}
  \clearpage
%}

\newpage
\begin{landscape}   % rotate only this page
\begin{singlespace}
\begin{figure}[htbp]
    \begin{minipage}{0.51\textwidth}
        \centering
        \includegraphics[width=\linewidth]{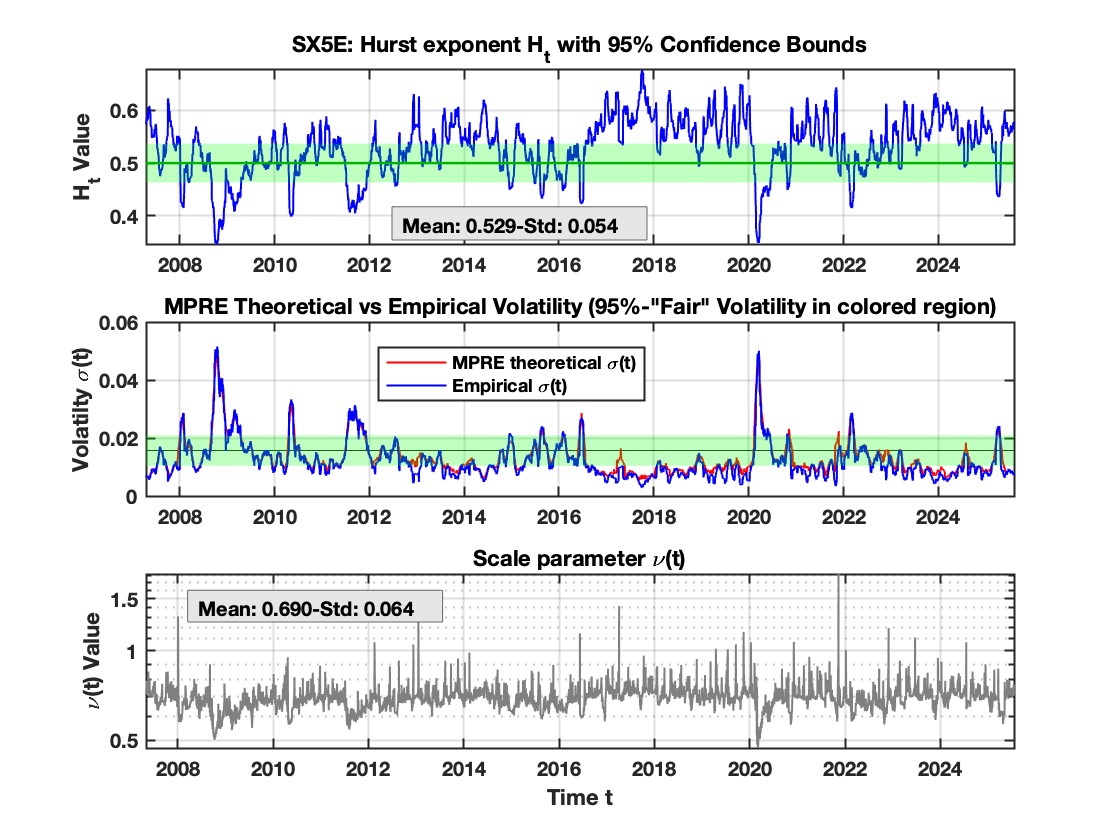}
        \caption{SX5E} \label{fig:SX5E}
    \end{minipage} 
    \begin{minipage}{0.51\textwidth}
        \centering
        \includegraphics[width=\linewidth]{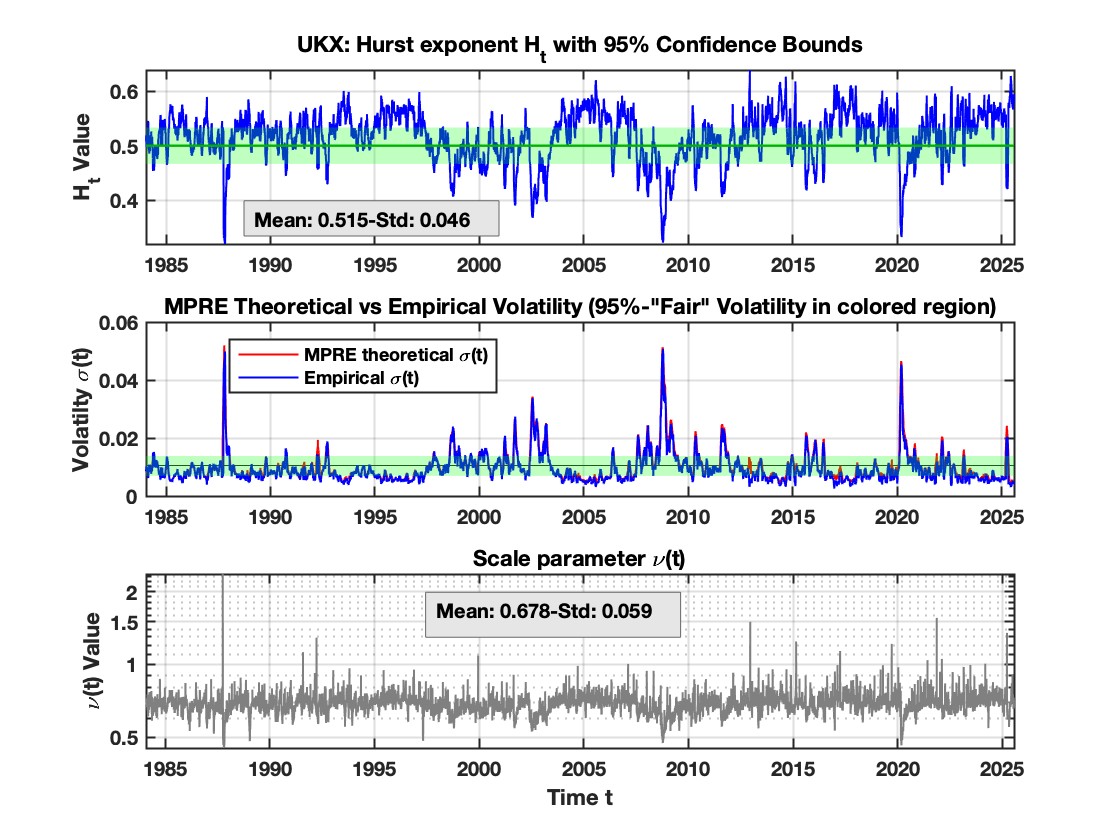}
        \caption{UKX Index} \label{fig:UKX}
    \end{minipage}
    \begin{minipage}{0.51\textwidth}
        \centering
        \includegraphics[width=\linewidth]{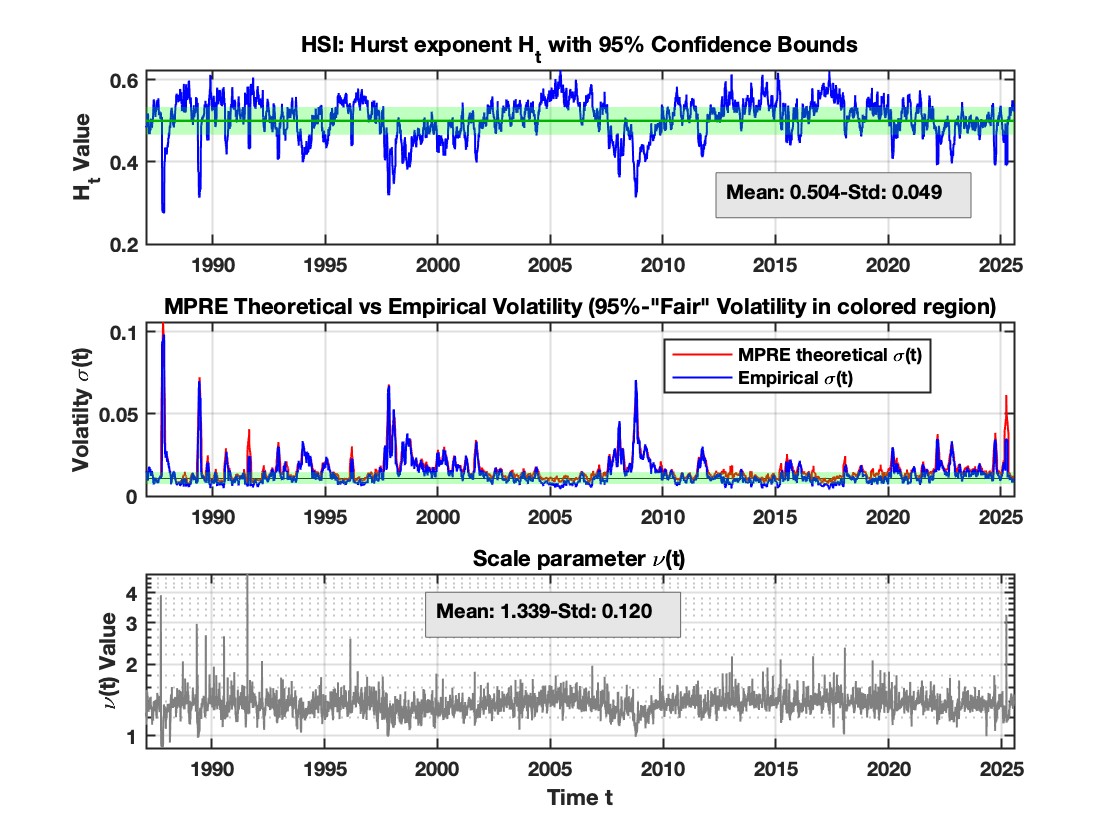}
        \caption{HSI Index} \label{fig:HSI}
    \end{minipage} \hfill
    \vspace{.5cm} \\ 
    \begin{minipage}{0.51\textwidth}
        \centering
        \includegraphics[width=\linewidth]{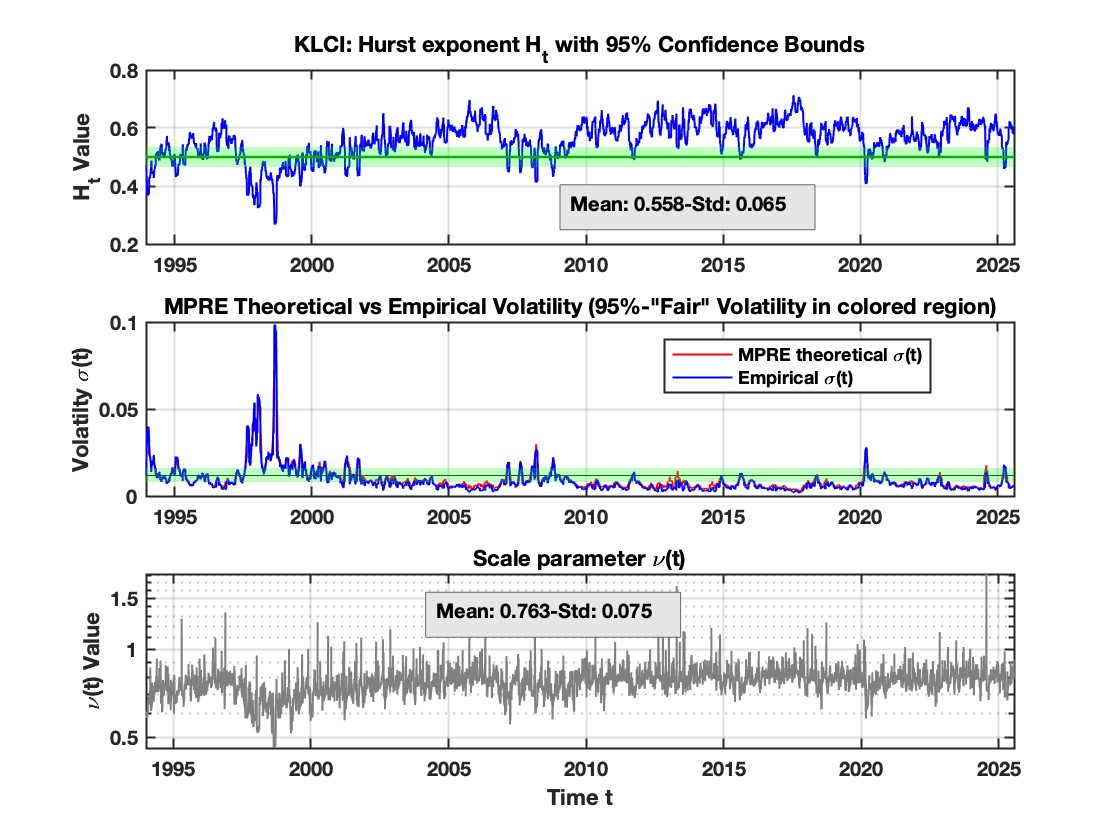}
        \caption{KLCI Index} \label{fig:KLCI}
    \end{minipage}
    \begin{minipage}{0.51\textwidth}
        \centering
        \includegraphics[width=\linewidth]{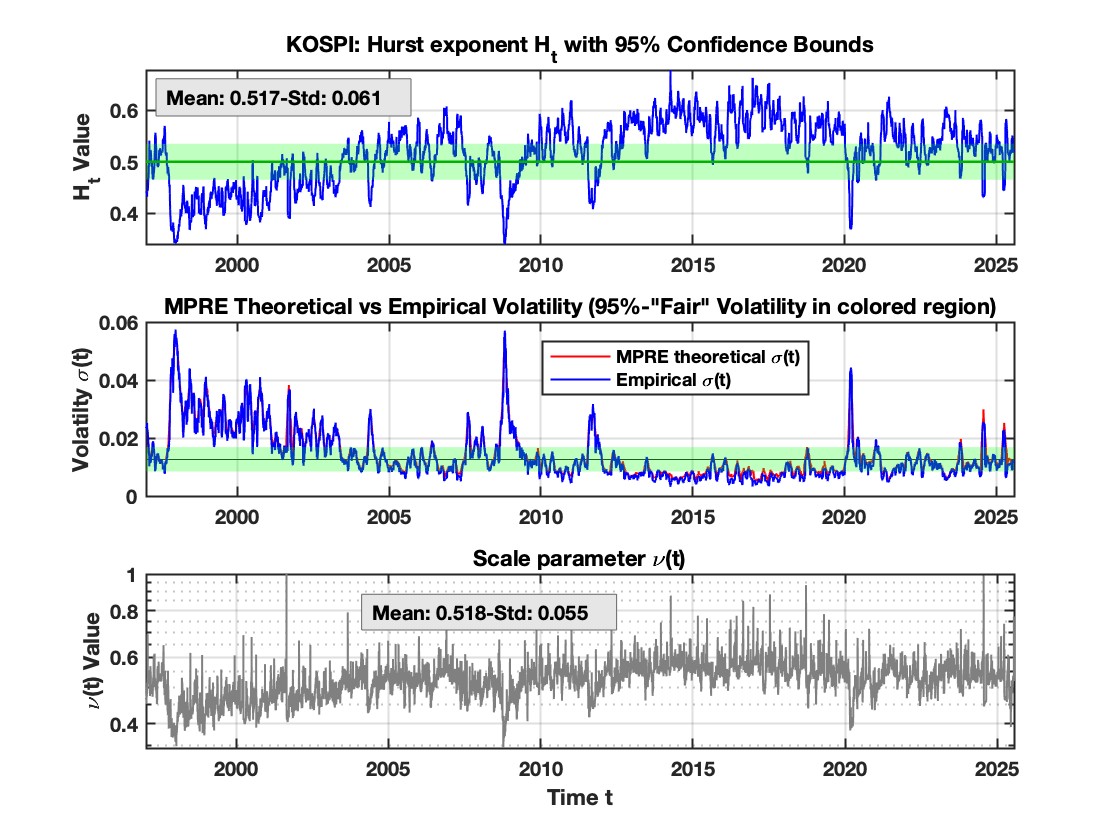}
        \caption{KOSPI Index} \label{fig:KOSPI}
    \end{minipage}
    \begin{minipage}{0.51\textwidth}
        \centering
        \includegraphics[width=\linewidth]{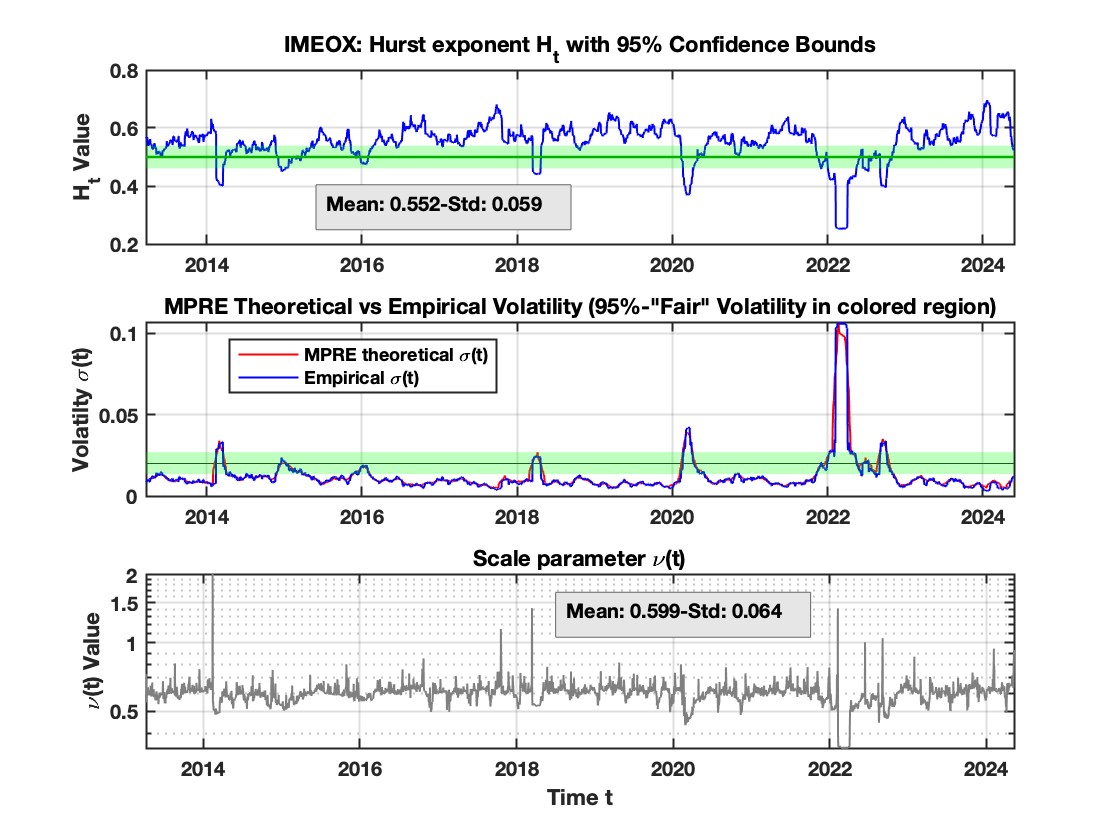}
        \caption{IMOEX Index} \label{fig:IMOEX}
    \end{minipage}
\end{figure}
\end{singlespace}
\end{landscape}

\newpage
\begin{landscape}   % rotate only this page
\begin{singlespace}
\begin{figure}[htbp]
    
    % Row 1
    \begin{minipage}{0.51\textwidth}
        \centering
        \includegraphics[width=\linewidth]{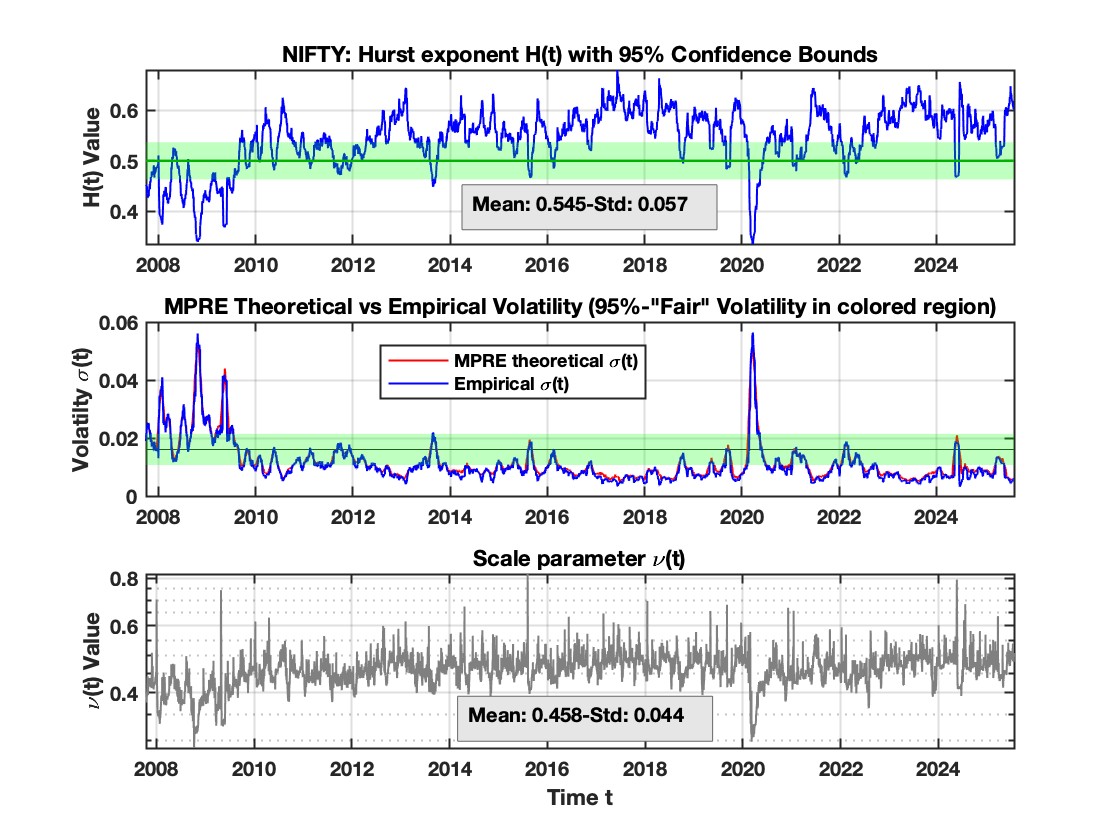}
        \caption{NIFTY Index} \label{fig:NIFTY}
    \end{minipage} 
    \begin{minipage}{0.51\textwidth}
        \centering
        \includegraphics[width=\linewidth]{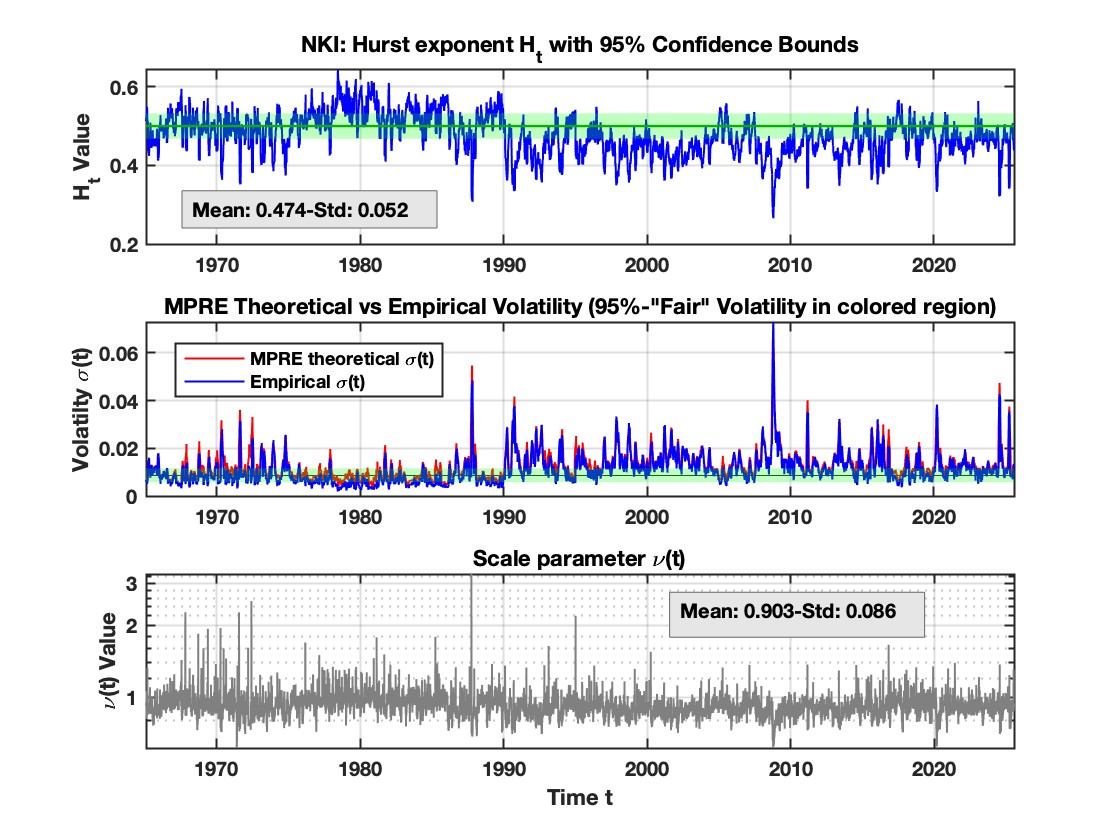}
        \caption{NKI Index} \label{fig:NKI}
    \end{minipage}
    \begin{minipage}{0.51\textwidth}
        \centering
        \includegraphics[width=\linewidth]{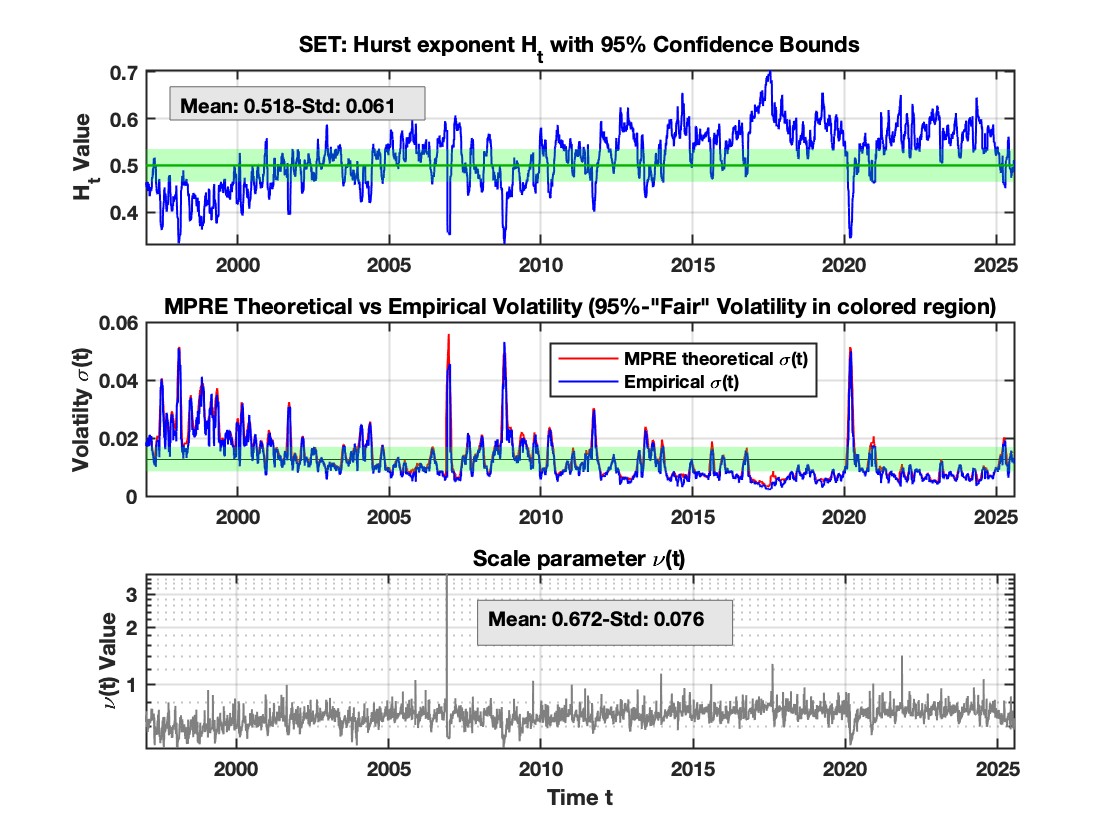}
        \caption{SET Index} \label{fig:SET}
    \end{minipage} \hfill
    \vspace{.5cm} \\ 
    
    % Row 2
    \begin{minipage}{0.51\textwidth}
        \centering
        \includegraphics[width=\linewidth]{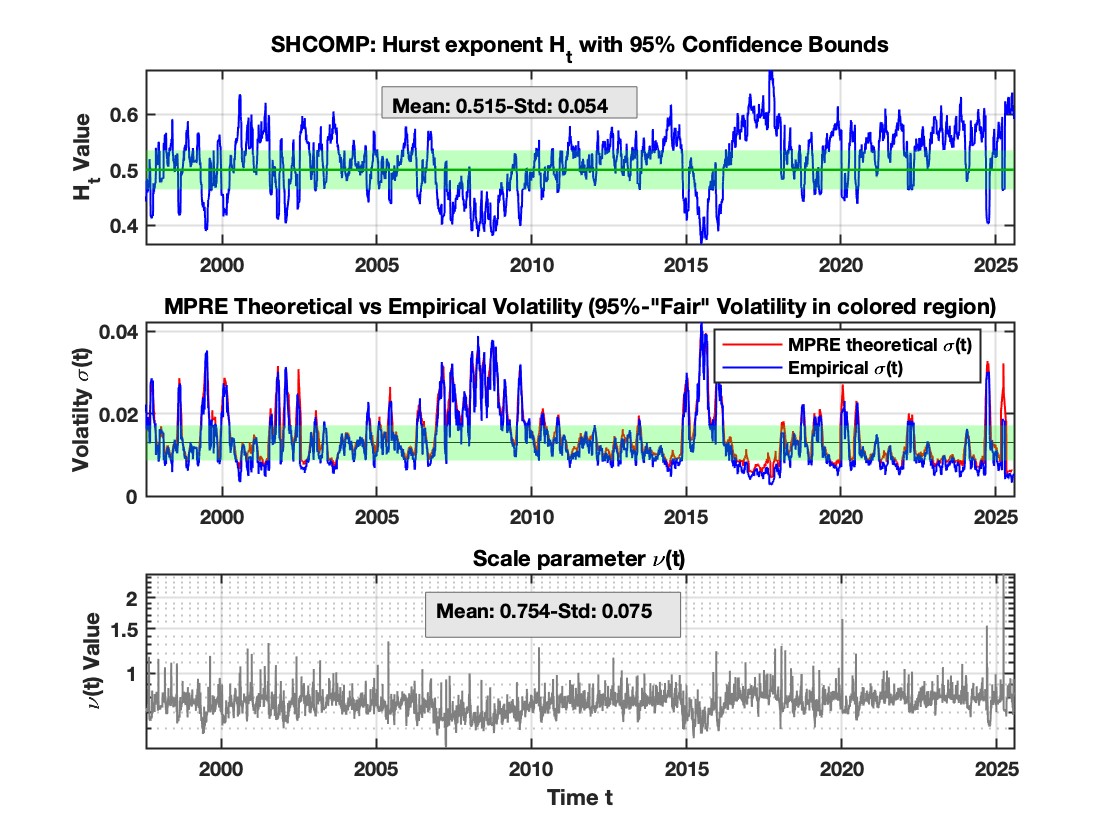}
        \caption{SHCOMP Index} \label{fig:SHCOMP}
    \end{minipage}
    \begin{minipage}{0.51\textwidth}
        \centering
        \includegraphics[width=\linewidth]{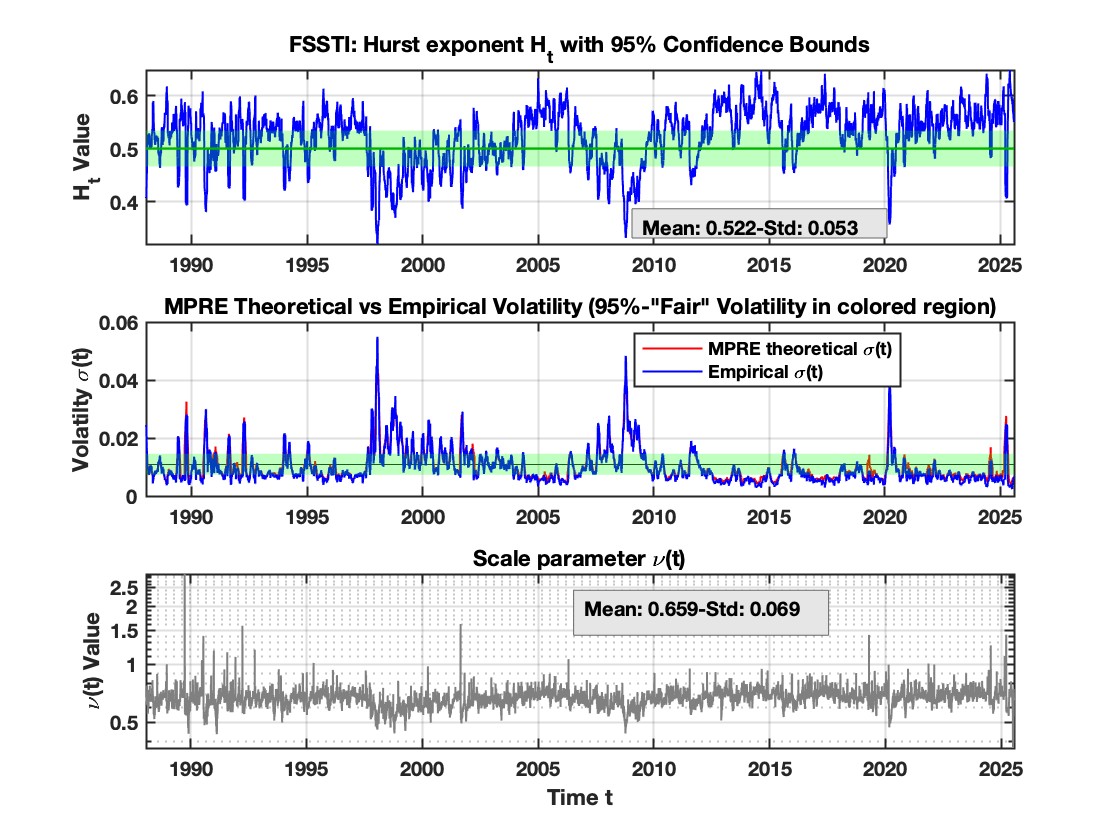}
        \caption{FSSTI Index} \label{fig:FSSTI}
    \end{minipage}
    \begin{minipage}{0.51\textwidth}
        \centering
        \includegraphics[width=\linewidth]{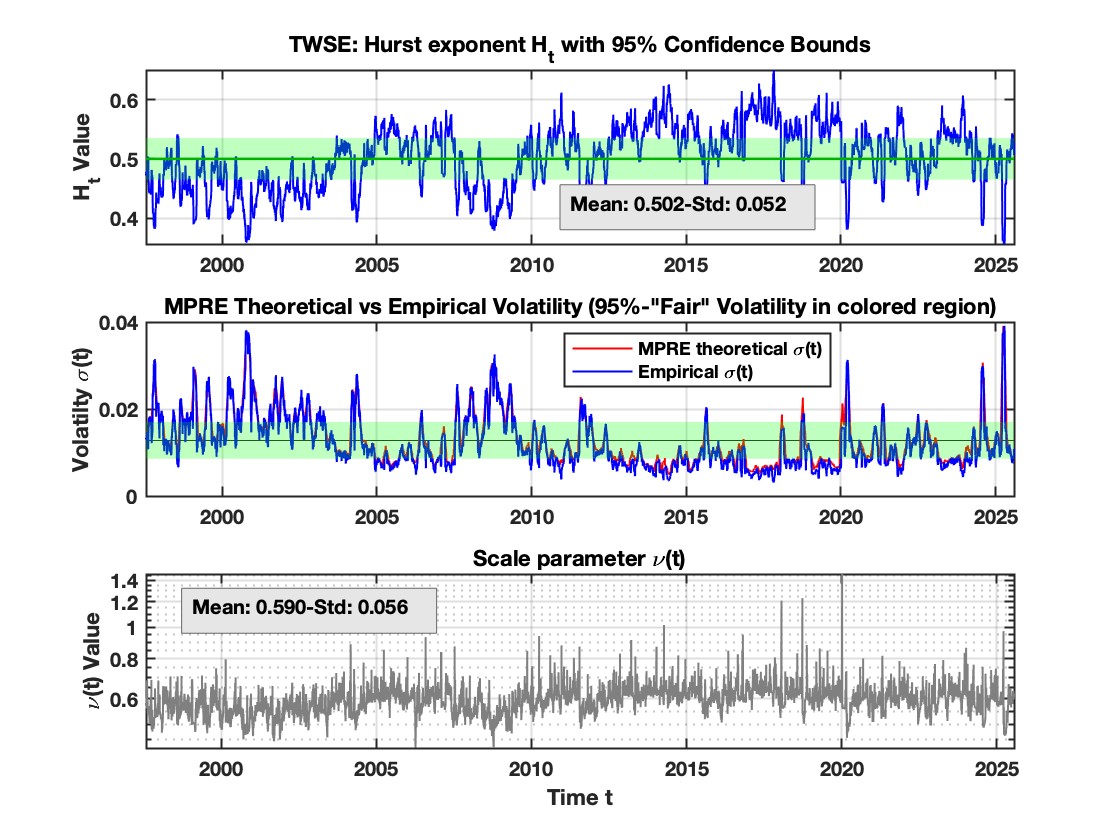}
        \caption{TWSE Index} \label{fig:TWSE}
    \end{minipage}
\end{figure}
\end{singlespace}
\end{landscape}

\subsection{Discussion of results}

The empirical analysis across a broad set of equity indices reveals several robust features in the time evolution of the estimated Hurst--H\"older exponent:

\begin{enumerate}[left=0pt, nosep, align=left]

\item The estimated sequences exhibit pronounced stability across markets, sample periods, and time-series lengths. Augmented Dickey--Fuller tests uniformly support trend stationarity, confirming that $\hat{\mathbf{H}}^{2,\delta,N}$ is stationary in all cases.

\item All indices display mean reversion toward the benchmark $H_t = \tfrac12$. For 11 of the 14 indices, the sample mean $\hat{\bar{\mathbf{H}}}^{2,\delta,N}$ lies within the corresponding $95\%$ confidence interval; the only exceptions are KLCI ($0.5579$), IMOEX ($0.5524$), and NIFTY ($0.5448$), the last one only marginally above the upper bound. Importantly, local deviations from $H_t=1/2$ should not be interpreted as evidence of arbitrage. Such deviations reflect short-lived statistical predictability in the second-order structure of increments, but do not imply the existence of predictable \emph{drift} components that could be exploited by admissible trading strategies. Market frictions, transaction costs, liquidity constraints, discretisation, and the distinction between $L^2$-forecastability and economic profitability further limit the tradability of such signals. Hence, $H_t\neq \tfrac12$ is best interpreted as \emph{informational disequilibrium} or \emph{transient inefficiency}---a temporary departure from local martingale behavior---consistent with the literature on near-efficiency, microstructure distortions, and the divergence between statistical and economic arbitrage in semimartingale models.

\item The roughness of the estimated paths, coupled with their persistent clustering around $H_t=\tfrac12$, is economically meaningful. It indicates that deviations from informational efficiency tend to be temporary, with markets displaying a systematic propensity to revert toward the efficient state. This pattern suggests the presence of continuously offsetting disequilibria.

\item The kurtosis of $\hat{\mathbf{H}}^{2,\delta,N}$ (ranging from 2.47 to 4.48, except for IMOEX at 7.33) and the small skewness (between $-0.91$ and $-0.13$, with IMOEX at $-1.38$) suggest that a normal distribution provides a reasonable approximation to the empirical distribution of the estimated exponent.

\item The slight left skewness of $\hat{\mathbf{H}}^{2,\delta,N}$ reflects the greater frequency of large negative shifts in $H_t$ relative to positive ones. These negative movements typically occur during market downturns and are consistent with the asymmetric response of market participants to news. Confidence building is generally gradual, producing small positive adjustments in $H_t$, whereas confidence erosion is often triggered instantaneously by adverse or uncertainty-increasing information.

\item Deviations of the Hurst--H\"older exponent from $1/2$ are sufficiently frequent and persistent to challenge the notion that arbitrage pressures instantaneously eliminate inefficiencies. Instead, all indices exhibit a systematic alternation between periods of efficiency and periods of inefficiency of either a \emph{positive} type ($H_t>1/2$, momentum) or a \emph{negative} type ($H_t<1/2$, reversal). This oscillatory behavior constitutes a robust empirical regularity, clearly visible in the magnitude and frequency of the deviations plotted in Figures~\ref{fig:SPX}--\ref{fig:TWSE}.
\end{enumerate}
\vspace{.5cm}

The patterns summarized above are consistent with and actually support the efforts to model stochastic volatility (as well as the $H_t$ process in MPRE) as a fractional Ornstein-Uhlenbeck processes \citep{Gatheral2018,AngeliniBianchi2023,Bianchi2023b}. The error metrics reported in Table \ref{tab:errors} confirm that the MPRE-based volatility reconstruction is systematically more robust than both GARCH(1,1) and HAR-RV, common benchmarks in literature. In almost all indices, the MPRE method achieves lower RMSE, MAE and SUP errors, indicating a superior ability to capture local dependence structures and adapt to rapid changes in market roughness. This empirical advantage validates the theoretical link established in Proposition 1 and demonstrates that modelling volatility through the time-varying Hurst-H\"{o}lder exponent provides a more stable and reliable approximation of true market variability.
\begin{table}
    \centering
    \scriptsize
    \caption{Forecasting errors for competing volatility models. RMSE is multiplied by $10^{6}$, MSE by $10^{3}$ and  SUP by $10^{2}$. The smallest errors are in bold.}
    \begin{tabular}{|l|ccc|ccc|ccc|}
    \hline
       Model  &  & GARCH(1,1) &  &  & HAR-RV &  &  & MPRE & \\
    \hline
        \diagbox{Index}{Error} & RMSE & MAE & SUP & RMSE & MAE & SUP & RMSE & MAE & SUP\\
        \hline
        SPX & \textbf{3.0086} & 1.1681 & \textbf{2.6519} & 9.2264 & 2.2672 & 4.0556 & 6.7183 & \textbf{1.0376} & 3.1390\\
        CCMP & 3.7064 & 1.3502 & 1.9865 & 7.7274 & 2.1100 & 2.1849 & \textbf{2.4265} & \textbf{1.2791} & \textbf{1.7247}\\
        SX5E & 6.5604 & 2.8259 & 2.0435 & 9.5245 & 2.3016 & 2.1104 & \textbf{4.3835} &  \textbf{0.9976}&\textbf{1.2676} \\
        UKX & 3.0889 & 1.2424 & 2.0229 & 6.5365 & 1.7930 & 3.3257 & \textbf{2.6835} & \textbf{0.9986} & \textbf{1.8743}\\
        HSI & 9.3564 & 1.9307 & \textbf{4.8500} & 32.6058 & 4.3230 & 5.3659 & \textbf{8.2346} & \textbf{1.8092} & 5.5939\\
        KLCI & 4.4110 & 1.1811 & 3.3904 & 31.1668 & 3.4145 & 7.9895 & \textbf{4.1960} & \textbf{1.1265} & \textbf{3.2146}\\
        KOSPI & 3.7453 & 1.4077 & 1.2206 & 8.8861 & 2.3963 & 1.4947 & \textbf{3.5473} & \textbf{0.9740} & \textbf{1.1011}\\
        IMOEX & 17.5413 & 2.0044 & 5.8044 & 65.6543 & 5.3347 & 8.4526 & \textbf{10.9599} & \textbf{1.3161} & \textbf{4.4875} \\
        NIFTY & 3.2598 & 1.2117 & \textbf{1.4163} & 8.3177 & 2.2043 & 1.8745 & \textbf{2.7999} & \textbf{1.0756} & 1.4225\\
        NKI & \textbf{5.9450} & \textbf{1.7055} & \textbf{2.3721} & 10.2265 & 2.3808 & 3.6390 & 9.1991 & 2.2139 & 3.0660\\
        SET & 6.0209 & 1.7973 & \textbf{2.1513} & 14.7149 & 2.9710 & 4.3644 & \textbf{5.5370} & \textbf{1.4114} & 3.2573\\
        SHCOMP & 4.5343 & 1.6631 & \textbf{1.0303} & 7.2269 & 2.2171 & 1.0777 & \textbf{1.9216} & \textbf{1.1472} & 1.2621 \\
        FSSTI & 6.1286 & 1.7082 & 2.3490 & 7.4662 & 2.0706 & 2.1069 & \textbf{3.5517} & \textbf{1.1994} & \textbf{1.8852}\\
        TWSE & \textbf{3.6117} & 1.4516 & 1.3995 & 8.2381 & 2.3171 & 1.5110 & 3.6352 & \textbf{1.3348} & \textbf{1.3746}\\
        \hline
    \end{tabular}
    \label{tab:errors}
\end{table}
\vspace{.1cm}\\

Within this interpretative framework and with reference to the stock price $S_t$, the notion of a \textit{fair} volatility at significance level $\alpha$ may be understood from equation \eqref{eq:confint} as the volatility 
\begin{equation*}
    \sigma(\alpha):=\text{sd}\left\{\ln\left(\frac{S_t}{S_{t-1}}\right):\hat{H}^{2,\delta,N}_{t}\in\left[\frac{1}{2} \mp z_{1-\alpha/2}\sqrt{\text{Var}(\hat{H}^{2,\delta,N}_t|_{H_t=1/2})}\right]\right\}.
\end{equation*}
where as usual $z_{1-\alpha/2}=\Phi^{-1}(1-\alpha/2)$ is the inverse cumulative distribution function (quantile function) of the standard normal distribution.\\
Data in Table \ref{tab:efficiency_metrics} show the percentage of the estimated Hurst exponents and the estimated volatility falling within the respective $95\%$ efficient confidence interval. A comprehensive examination of how these metrics might be employed to assess degrees of market efficiency remains a valuable avenue for future research. The present study, therefore, focuses on providing the foundational data, leaving its detailed interpretation for subsequent analysis.\\
%\begin{table}[htbp]
%\centering
%\scriptsize 
%\caption{Market Efficiency Metrics by Index. Percentage of time the Hurst exponent $H_t$ is within the $95\%$ efficient confidence interval (column 2) and the volatility $\sigma(t)$ is falls within the $95\%$ confidence band.}\label{tab:efficiency_metrics}
%\begin{tabular}{@{} lcc @{}}
%\toprule
%\textbf{Index} & $\% \sep H_t\in 95\%\text{-CI}$  & $\% \sep \sigma(t)\in 95\%\text{-CI}$ \\
%\midrule
%SPX    & 43.62 & 50.65 \\
%CCMP   & 32.36 & 46.37 \\
%SX5E   & 36.14 & 41.02 \\
%UKX    & 56.01 & 53.31 \\
%HSI    & 37.08 & 59.52 \\
%KLCI   & 23.74 & 28.15 \\
%KOSPI  & 44.10 & 45.35 \\
%IMOEX  & 61.17 & 14.23 \\
%NIFTY  & 51.51 & 27.56 \\
%NKY    & 46.75 & 47.43 \\
%SET    & 51.85 & 56.21 \\
%SHCOMP & 33.07 & 50.72 \\
%FSSTI  & 39.09 & 41.74 \\
%TWSE   & 46.00 & 45.21 \\
%\bottomrule
%\end{tabular}
%\end{table}

\begin{table}[htbp]
\centering
\scriptsize
\scriptsize 
\caption{Market Efficiency Metrics by Index. Percentage of time the estimated Hurst exponent and volatility are within the corresponding $95\%$ efficient confidence interval.}\label{tab:efficiency_metrics}
\begin{tabular}{@{} lccccccc @{}}
\toprule
\textbf{Metric / Index} & \textbf{SPX} & \textbf{CCMP} & \textbf{SX5E} & \textbf{UKX} & \textbf{HSI} & \textbf{KLCI} & \textbf{KOSPI} \\
\midrule
$\text{Pct} \quad \hat{H}^{2,\delta,N}_t \in 95\%\text{-CI} $ & 47.79 & 40.51 & 40.44 & 48.70 & 51.71 & 21.24 & 35.68 \\
$\text{Pct} \quad \hat{\sigma}_t\in 95\%\text{-CI}$ & 51.03 & 46.94 & 40.92 & 53.78 & 59.98 & 28.14 & 45.32 \\
 & & & & & & &\\
\midrule
 & \textbf{IMOEX} & \textbf{NIFTY} & \textbf{NKI} & \textbf{SET} & \textbf{SHCOMP} & \textbf{FSSTI} & \textbf{TWSE} \\
\midrule
$\text{Pct} \quad \hat{H}^{2,\delta,N}_t \in 95\%\text{-CI}$ & 23.66 & 28.79 & 41.13 & 37.19 & 43.14 & 39.09 & 43.22 \\
$\text{Pct} \quad \hat{\sigma}_t\in 95\%\text{-CI}$ & 14.23 & 14.42 & 47.95 & 41.54 & 51.26 & 42.06 & 45.29 \\
\bottomrule
\end{tabular}
\end{table}

The middle panels of Figures \ref{fig:SPX}–\ref{fig:TWSE} offer empirical evidence that Proposition \ref{prop:1} holds. In all cases examined, the theoretical volatility determined using formula \eqref{eq:IncrVar} is comparable to that estimated empirically, with the exceptions represented by outliers, where theoretical volatility is systematically higher than actual volatility. In addition to suggesting that the MPRE is definitely a good model for price dynamics in financial markets, the theoretical relationship identified allows us to respond to the primary requirement of this work, namely to link volatility to the time-varying dependence observed in the paths followed by prices (or indices in this case). The empirical validation of equation \eqref{eq:IncrVar} establishes a foundation for deriving a fair volatility benchmark for any financial time series. Consequently, this framework provides a rigorous, model-based methodology to address a long-standing practical challenge: determining whether observed volatility is anomalously high or low at any point in time. This approach moves beyond traditional reliance on practitioner heuristics, ex-post portfolio analysis, or comparative measures like implied volatility. Instead, it enables a theoretically grounded assessment by quantifying the divergence between realized volatility and its efficient-market benchmark—the level expected if $H_t$ were precisely $1/2$. This benchmark represents an equilibrium state toward which market prices gravitate to correct transient inefficiencies.
The direction and magnitude of deviation from this equilibrium are economically informative. Significantly elevated volatility signals market roughness and negative autocorrelation—a state of heightened irregularity that typically triggers rapid mean-reversion. Conversely, suppressed volatility indicates excessive smoothness and positive autocorrelation, a condition of persistent regularity that often endures over longer horizons.
This asymmetry in market correction mechanisms finds a compelling explanation in the principles of behavioral finance. The cognitive and emotional biases detailed in Table \ref{tab:FinIn}—such as overreaction, herding, and loss aversion—provide a microstructure rationale for why markets correct disruptive volatility spikes more swiftly than they resolve periods of trending, but calmer, inefficiency.

\section{Conclusion and further developments} \label{sec:Conclusion}
This study establishes the Hurst--H\"older exponent as an informationally equivalent yet conceptually superior alternative to volatility for measuring financial risk under a very general and comprehensive local fractional price dynamics. Its use offers three main advantages. First, \emph{path-dependent risk}: the exponent quantifies local roughness, capturing deviations from semimartingale behaviour that volatility alone cannot, thereby characterizing not only the magnitude but also the \emph{form} of randomness. Second, \emph{absolute benchmarking}: unlike volatility---a purely relative measure---the exponent possesses an intrinsic scale anchored at the martingale benchmark $H_t=\tfrac12$. Third, \emph{theoretical synthesis}: fluctuations of $H_t$ around $\tfrac12$ provide a unified representation in which market efficiency and behavioural deviations correspond to alternating regimes rather than competing paradigms.

Following \citep{Lobodaetal2021}, Proposition~\ref{prop:1} shows that the Hurst--H\"older exponent can be mapped into realized volatility, enabling the construction of confidence intervals for the volatility level implied by informational efficiency. This motivates the notion of \emph{fair volatility}, defined as the volatility consistent with $H_t=\tfrac12$. Volatility inferred from the exponent thus acquires an absolute interpretation: instead of being assessed only relative to its past, it becomes an indicator of the market's distance from theoretical equilibrium.

The mean-reverting behaviour of the estimated exponent further implies that volatility derived from this framework is informative about the correction required for the market to return to equilibrium following a deviation. A more detailed analysis of the mean-reversion dynamics may allow estimation of the expected time to re-equilibrate.

The proposed framework paves the way for several methodological extensions. Developing a joint estimation procedure for $(H_t,\nu_t)$ would clarify their respective roles in volatility formation and reduce identification concerns. Additional Monte Carlo experiments incorporating jumps, heavy tails, and leverage effects would improve understanding of estimator robustness in more realistic environments. On the empirical side, dependence-aware inference and out-of-sample forecasting---benchmarked against HAR--RV, GARCH, and rough-volatility models---represent natural next steps. Assessing the economic relevance of fair volatility, for instance for portfolio scaling or risk management, would further ground the concept in applied finance. Extending the analysis across horizons, asset classes, and incorporating option-implied information could broaden applicability and deepen the empirical foundations of the approach. Finally, an additional natural direction for future research involves extending the proposed fair volatility framework to high-frequency intraday data. The primary challenge in this environment is the pervasive presence of market microstructure noise—such as bid-ask bounce and discrete pricing—which mechanically biases regularity estimators downward. This noise generates a spurious roughness that can easily conflate mechanical trading frictions with genuine informational inefficiency. Future work should therefore focus on developing dynamically optimized sampling protocols that balance the preservation of large intraday sample sizes with the necessary filtering of microstructural distortions, ultimately paving the way for the real-time, intraday monitoring of martingale-consistent fair volatility.
\\

\vspace{7pt}
\noindent \footnotesize{\textbf{Funding}. This research was supported by Sapienza University of Rome under Grant No. RM120172B346C021.}
\vspace{7pt}\\
\noindent \textbf{Acknowledgments}
The Authors wish to thank the anonymous Reviewers for their valuable comments and suggestions which have helped improve the quality of the work.

\bibliographystyle{plain} 
\bibliography{Bibliography10}

\end{document}